\documentclass[12pt]{article}
\usepackage[top=1in, bottom=1in, left=1in, right=1in]{geometry}
\usepackage{setspace}
\usepackage[T1]{fontenc}
\usepackage{times}
\usepackage{booktabs}
\usepackage{rotating}
\usepackage{graphicx}
\usepackage[section]{placeins} %Placeins.sty keeps floats `in their place', preventing them from floating past a "\FloatBarrier" command into another section.  To use it, declare "\usepackage{placeins}" and insert "\FloatBarrier" at places that floats should not move past, perhaps at every "\section".  
\usepackage[small, bf, tableposition=top]{caption}
\usepackage{subcaption}
\usepackage{pdfpages}
\usepackage{pdflscape}
\usepackage{textcomp}
\usepackage{longtable}
\usepackage{nicefrac}
\usepackage{adjustbox}	% to adjust the size of objects to fit into a page
\usepackage[hyphens]{url}
\usepackage{bbm}
\usepackage{cancel}
\usepackage{natbib}
\usepackage{bibunits}
\bibpunct{(}{)}{;}{a}{,}{,}
\usepackage{array}
\usepackage{floatrow}
\usepackage{float}
\floatstyle{plaintop}
\restylefloat{figure}

\usepackage{amsmath, amssymb, amsthm}
\usepackage{mathtools}
\usepackage[pdftex]{hyperref}
\hypersetup{
    colorlinks=true,
    linkcolor=red,
    citecolor=blue,
    filecolor=magenta,      
    urlcolor=cyan,
}

\theoremstyle{plain}
\newtheorem{theorem}{Theorem}

\newtheorem{corollary}{Corollary}

\theoremstyle{plain}
\newtheorem{definition}{Definition}
\newtheorem{assumption}{Assumption}
\newtheorem{remark}{Remark}

\theoremstyle{definition}
\newtheorem{example}{Example}
\newtheorem*{example*}{Example}

\usepackage{titlesec}
\titlespacing*{\section}{0pt}{1.5ex plus 1ex minus .2ex}{0.8ex plus .2ex}
\titlespacing*{\subsection}{0pt}{1.2ex plus 1ex minus .2ex}{0.8ex plus .2ex}

\newcommand{\indep}{\perp \!\!\!\perp}

\newcommand{\cW}{\mathcal{W}}

\usepackage{dcolumn}
\newcolumntype{d}[0]{D{.}{.}{5}}

\usepackage{graphicx}
\usepackage[space]{grffile}
\usepackage{multirow}

\title{When do common time series estimands  \\
have nonparametric causal meaning?\footnote{This is a revised version of ``A nonparametric dynamic causal model for macroeconometrics.'' 
We thank Iavor Bojinov, Gary Chamberlain, Fabrizia Mealli, James M. Robins, Frank Schorfheide and James H. Stock for conversations that have helped developed our thinking on causality. 
We are also grateful to many others, including numerous conference participants, for their comments on earlier versions of this paper. Rambachan gratefully acknowledges financial support from the NSF Graduate Research Fellowship under Grant DGE1745303. All remaining errors are our own.}}
\author{Ashesh Rambachan\thanks{MIT, Department of Economics: \texttt{asheshr@mit.edu}} \\
\and Neil Shephard\thanks{Harvard University, Department of Economics and Department of Statistics: \texttt{shephard@fas.harvard.edu}}
}

\begin{document}

\maketitle
\begin{abstract} 
In this paper, we introduce the direct potential outcome system as a framework for analyzing dynamic causal effects of assignments on outcomes in observational time series settings.
We provide conditions under which common predictive time series estimands, such as the impulse response function, generalized impulse response function, local projection, and local projection instrumental variables, have a nonparametric causal interpretation in terms of dynamic causal effects. 
The direct potential outcome system therefore provides a foundation for analyzing popular reduced-form methods for estimating the causal effect of macroeconomic shocks on outcomes in time series settings. 
\end{abstract}

%%%%%%%%%%%%%%%%
% Introduction %
%%%%%%%%%%%%%%%%
\section{Introduction}
In this paper, we introduce the \textit{direct potential outcome system} as a framework for analyzing dynamic causal effects of assignments on outcomes in observational time series settings.
We consider settings in which there is a single unit (e.g., macroeconomy or market) observed over time. 
At each time period $t \geq 1$, the unit receives a vector of assignments $W_{t}$, and an associated vector of outcomes $Y_{t}$ are generated. 
The outcomes are causally related to the assignments through a \textit{potential outcome process}, which is a stochastic process that describes what would be observed along counterfactual assignment paths.
A dynamic causal effect compares the potential outcome process along different assignment paths at a fixed point in time. 

Importantly, we place no functional form restrictions on the potential outcome process, no restrictions on the extent to which past assignments may causally affect outcomes, nor common time series assumptions such as ``invertibility'' or ``recoverability'' on the system of potential outcomes and assignments. 
Most leading econometric models used to study dynamic causal effects in time series settings, such as the structural vector moving average model and (both linear and non-linear) structural vector autoregressions, can be cast as special cases of the direct potential outcome system by introducing these assumptions on the potential outcome process.
In this sense, the direct potential outcome system provides a flexible foundation upon which to analyze dynamic causal effects in time series settings.

We analyze conditions under which reduced-form time series estimands, such as the impulse response function \citep{Sims(80)}, generalized impulse response function \citep{KoopPesaranPotter(96)}, local projection \citep{Jorda(05)} and local projection instrumental variables \citep{JordaSchularickTaylor(15)}, have a nonparametric causal interpretation in terms of dynamic causal effects of assignments on outcomes. 
Under what conditions do these common time series estimands have a nonparametric causal interpretation as measuring how movements in the outcomes $Y_{t+h}$ for some $h \geq 0$ are caused by changes in the assignments $W_{t}$?
In exploring this question, we focus on four data environments, which place alternative assumptions on what output is observed by the researcher.

First, we analyze a benchmark case in which the researcher observes both the outcomes $Y_{t+h}$ and the assignments $W_{t}$ generated by the potential outcome system. 
We show that impulse response functions, local projections, and generalized impulse response functions of the outcome $Y_{t+h}$ on the assignment $W_{t}$ identify a dynamic average treatment effect, a weighted average of marginal average treatment effects, and a filtered average treatment effect respectively if the assignments $W_{t}$ are randomly assigned.
Random assignment requires that the assignment must be independent of the potential outcome process \textit{and} the assignments must be independent over time. 
This provides a new perspective on an influential literature in empirical macroeconomics that constructs measures of underlying economic shocks, and then uses these constructed measures to estimate dynamic causal effects using reduced form methods. 
\cite{NakamuraSteinsson(18)-JEP} refers to this empirical strategy as ``direct causal inference.''
Our first set of results therefore provide conditions that these constructed shocks must satisfy in order for the resulting reduced form estimates to be causally interpretable.

Second, we provide a special case of the direct potential outcome system to incorporate instrumental variables $Z_{t}$ that causally affect the assignment $W_{t}$ but not the outcome $Y_{t+h}$. 
Provided the researcher directly observes the instrument, the assignment, and outcome, it is natural to consider the causal interpretation of the time-series analogue of common instrumental variable estimands. 
We focus attention on dynamic instrumental variable estimands that take the ratio of an impulse response function of the outcome $Y_{t+h}$ on the instrument $Z_t$ (a ``reduced-form'') relative to an impulse response function of the assignment $W_{t+h}$ on the instrument $Z_{t}$ (a ``first stage''). 
We show that such dynamic instrumental variable estimands identify an appropriately defined dynamic ``local average treatment effect'' in the spirit of \cite{ImbensAngrist(94)}, provided the instrument is randomly assigned and satisfies a familiar monotonicity condition. 
The dynamic local average treatment effects that we characterize measure the average dynamic treatment effect of the assignment on the $h$-period ahead outcome conditional on the event that the instrument causally affects the treatment. 

We further analyze the case in which the researcher only observes the instrument $Z_{t}$ and outcome $Y_{t+h}$ but not the assignment $W_{t}$ itself. 
This is an important case.
Empirical researchers in macroeconomics increasingly use ``external instruments'' to identify the dynamic causal effects of \textit{unobservable} economic shocks on macroeconomic outcomes \citep[e.g., see][]{JordaSchularickTaylor(15), GertlerKaradi(15), NakamuraSteinsson(18)-QJE, RameyZubairy(18), StockWatson(18), PlagborgMollerWolf(20), JordaSchularickTaylor(20)}.
In this line of research, it is common for empirical researchers to analyze estimands that involve two distinct elements of the outcome vector $Y_{j,t+h}, Y_{k,t}$ and the instrument. 
For example, given an external instrument $Z_{t}$ for the unobserved monetary policy shock $W_{t}$ \citep[e.g.,][]{Kuttner(01), CochranePiazessi(02), GertlerKaradi(15)}, it is common to measure the dynamic causal effect of the monetary policy shock on unemployment $Y_{j,t+h}$ by (1) estimating a reduced-form impulse response function of unemployment on the external instrument, (2) estimating a ``first stage'' impulse response function of the Federal Funds rate $Y_{k,t}$ on the external instrument, (3) reporting the ratio of these impulse responses \citep[e.g.,][]{JordaSchularickTaylor(15), StockWatson(18), JordaSchularickTaylor(20)}.

We show that these dynamic IV estimands constructed are also causally interpretable, and nonparametrically identify a \textit{relative}, dynamic local average treatment effect that measures the causal response of the $h$-step ahead outcome $Y_{j,t+h}$ to a change in the treatment $W_{k,t}$ that raises the contemporaneous outcome $Y_{k,t}$ by one unit among compliers (i.e., in the monetary policy example, the dynamic causal effect of unemployment to a monetary policy shock that raises the Federal funds rate by one unit).
This result provides a motivation for the recent interest in external instruments in empirical macroeconomics. 
Provided one exists, an external instrument can identify causally interpretable estimands without resorting to functional form restrictions on the outcomes and without even directly observing the treatment itself.

Finally, we conclude by briefly discussing the most challenging data environment in which the researcher only observes the outcomes $Y_{t+h}$, but not the assignments $W_{t}$ nor any external instruments $Z_{t}$. 
This case is considered by much foundational work on model-based approaches to analyzing dynamic causal effects \citep[e.g.,][]{Sims(72), Sims(80)}. 
We consider this setting in order to place the direct potential outcome system in this broader context, and illustrate that researchers can recover familiar model-based approaches by introducing further assumptions on the potential outcome process.

Taking a step back, quantifying dynamic causal effects is one of the great themes of the broader time series literature. Researchers use a variety of methods such as Granger causality \citep{Wiener(56), Granger(69),WhiteLu(10)}, highly structured models such as DSGE models \citep{HerbstSchorfheide(15)}, state space modeling \citep{HarveyDurbin(86), Harvey(96),BrodersenEtAl(15),MenchettiBojinov(21)} as well as intervention analysis \citep{BoxTiao(75)} and regression discontinuity \citep{KuersteinerEtAl(18)}.  The nonparametric potential outcome framework we develop is distinct.  References to some of the more closely related work will be given in the next section.  This paper is not focused on estimators and the associated distribution theory: we do not have much to say in that regard which is novel. 
\citet{KolesarPlagborgMoller(24)} build on our framework to further characterize the dynamic causal estimands identified by local projections and local projections IV under weaker regularity conditions.
Importantly, \citet[][]{KolesarPlagborgMoller(24)} further describe how researchers can empirically estimate the weighting functions in our characterizations; reporting the estimated weighting function is a useful diagnostic for understanding the nonparametric causal estimand associated with these methods in any application.
\citet[][]{CaravelloMartinezBruera(24)} and \citet[][]{CasiniMccloskey(24)} also build on our framework to study the causal interpretation of non-linear local projections and high-frequency event studies respectively.
Finally, \citet[][]{ballinari2024semiparametricinferenceimpulseresponse} build on our identification results to propose double/debiased machine learning estimators for dynamic causal effects in observational time series.

\vspace{-1em}
\paragraph{Notation:} For a time series $\{A_t\}_{t \geq 1}$ with $A_t \in \mathcal{A}$ for all $t \geq 1$, let $A_{1:t} := (A_1, \hdots, A_t)$ and $\mathcal{A}^{t} := \bigtimes_{s=1}^{t} \mathcal{A}$.  $A \indep B$ says that random variables $A$ and $B$ are probabilistically independent.  

\section{The Direct Potential Outcome System and Dynamic Causal Effects}\label{sect:POS}
We now introduce the \textit{direct potential outcome system}, which extends the design-based approach developed in \cite{BojinovShephard(19)} to stochastic processes. We define a large class of casual estimands that summarize the dynamic causal effects of varying the assignment on future outcomes. As an illustration, we show that the direct potential outcome system recovers most leading structural models in macroeconometrics as special cases.

The nonparametric potential outcome framework we develop relates to a vast literature on dynamic treatment effects in small-$T$, large-$N$ panels. 
The panel work of \cite{Robins(86)} and \cite{AbbringHeckman(07)}, amongst others, led to an enormous literature on dynamic causal effects in panel data \citep{MurphyEtAl(01), Murphy(03),HeckmanNavarro(07), Lechner(11), HeckmanEtAl(16),BoruvkaAlmirallWitkiwitzMurphy(17),LuSuWhite(17),BlackwellGlynn(18),HernanRobins(18),BojinovRambachanShephard(21),MastakouriScholkopfJanzing(21)}.  
The direct potential outcome system is also related to \cite{AngristKuersteiner(11)} and \cite{AngristJordaKuersteiner(18)}. We discuss their work in Section \ref{section: links to macro}. 

\subsection{The Direct Potential Outcome System}
There is a single unit. At each time period $t \geq 1$, the unit receives a $d_w$-dimensional assignment $\{W_t\}_{t \geq 1}$. Associated with this \textit{assignment process}, we observe a $d_y$-dimensional outcome $\{Y_t\}_{t \geq 1}$. 
The outcomes are causally related to the assignments through the potential outcome process, which describes what outcome would be observed at time $t$ along a particular path of assignments.

\begin{assumption}[Assignment and Potential Outcome]\label{assumption: underlying processes}
    The assignment process $\{W_{t}\}_{t \geq 1}$ satisfies $W_t \in \mathcal{W} := \bigtimes_{k=1}^{d_w} \mathcal{W}_{k} \subseteq \mathcal{R}^{d_w}$. The potential outcome process is, for any deterministic sequence $\{w_s\}_{s \geq 1}$ with $w_s \in \mathcal{W}$ for all $s \geq 1$, 
        $
            \left\{ Y_t(\{w_s\}_{s \geq 1}) \right\}_{t \geq 1},
        $
    where the time-$t$ potential outcome satisfies $Y_{t}(\{w_s\}_{s \geq 1}) \in \mathcal{Y} \subseteq \mathbb{R}^{d_y}$. 
\end{assumption}

% \noindent The simplest case is when the assignment is scalar and binary $\mathcal{W} = \{0, 1\}$, in which case $W_t=1$ corresponds to ``treatment'' and $W_t=0$ is ``control.'' 

The potential outcome $Y_t(\{w_s\}_{s \geq 1})$ may depend on future assignments $\{w_s\}_{s \geq t + 1}$. Our next assumption rules out this dependence, restricting the potential outcome to only depend on past and contemporaneous assignments. % \footnote{Such a restriction is one of the nine \cite{BradfordHill(65)} criteria for causality (``temporality'').}

\begin{assumption}[Non-anticipating Potential Outcomes]\label{assumption: non-anticipating potential outcomes}
    For each $t \geq 1$, and all deterministic $\{w_t\}_{t \geq 1}, \{w_t^\prime\}_{t \geq 1}$ with $w_t, w_t^\prime \in \mathcal{W}$, $Y_t(w_{1:t}, \{w_{s}\}_{s \geq t + 1}) = Y_t(w_{1:t}, \{w_{s}^\prime\}_{s \geq t + 1})$ almost surely.
\end{assumption}

\noindent Assumption \ref{assumption: non-anticipating potential outcomes} is a stochastic process analogue of non-interference \citep{Cox(58), Rubin(80)}, extending \cite{WhiteKennedy(09)} and \cite{BojinovShephard(19)}. 
It still allows for rich dependence on past and contemporaneous assignments.
Under Assumption \ref{assumption: non-anticipating potential outcomes}, we drop references to the future assignments in the potential outcome process, and write $\{Y_t(\{w_s\}_{s \geq 1})\}_{t \geq 1} = \{Y_t(w_{1:t})\}_{t \geq 1}$.
The set $\left\{ Y_t(w_{1:t}) \colon w_{1:t} \in \mathcal{W}^{t} \right\}$ collects all the potential outcomes at time $t$. 

Together, the assignments and potential outcome generate the output of the system. 
\begin{assumption}[Output]\label{asm: fully observable output}
    The output is $\{W_t, Y_t\}_{t \geq 1} = \{W_t, Y_t(W_{1:t})\}_{t \geq 1}$, where $\{ Y_t\}_{t \ge 1}$ is the outcome process.  
\end{assumption}
% \noindent The outcome process is the potential outcome process evaluated at the assignment process. 

%Even if both the assignment and outcome processes are observed, so we face the usual causal challenge. We observe the potential outcomes associated with the realized assignment path, but do not observe the potential outcomes associated with counterfactual assignment paths.

Finally, we assume that the assignment process is sequentially probabilistic, meaning that any assignment vector may be realized with positive probability at time $t$ given the history of the observable stochastic processes up to time $t-1$. Let $\{\mathcal{F}_{t}\}_{t \geq 1}$ denote the natural filtration generated by (the $\sigma$-algebra of) the realized $\{w_t, y_t\}_{t \geq 1}$. 

\begin{assumption}[Sequentially probabilistic assignment process]\label{asm: sequentially probabilistic treatment}
    The assignment process satisfies $0 < P(W_t = w \mid \mathcal{F}_{t-1}) < 1$ with probability one for all $w \in \cW$. Here the probabilities are determined by a filtered probability space of $\{W_t,\{Y_t(w_{1:t}),w_{1:t} \in \mathcal{W}^t\}\}_{t\ge 1}$.
\end{assumption}

\noindent This is the time series analogue of the ``overlap'' condition in cross-sectional causal studies. We make this assumption throughout the paper in order to focus attention on the causal interpretation of common time series estimands in the presence of rich dynamic causal effects. 
Understanding how violations of Assumption \ref{asm: sequentially probabilistic treatment} affect the causal interpretation and estimation of common time series estimands is an important but separate issue.

By putting these assumptions together, we define a direct potential outcome system. 
\begin{definition}[Direct Potential Outcome System]\label{defn: potential outcome system}
    Any $\{W_t, \{Y_t(w_{1:t}) \colon w_{1:t} \in \mathcal{W}^{t}\}\}_{t \geq 1}$ satisfying Assumptions \ref{assumption: underlying processes}-\ref{asm: sequentially probabilistic treatment} is a \textbf{direct potential outcome system}. 
\end{definition}

\noindent We refer to Definition \ref{defn: potential outcome system} as a ``direct'' potential outcome system in order to emphasize that it focuses on nonparametrically modelling the direct causal effects of the assignment process $\{W_{t}\}$ on the outcomes $\{Y_{t}\}$.
We do not, however, explicitly allow for the assignment $W_{t}$ to have a causal effect on future assignments $W_{s}$ for $s > t$. That is, we do not introduce a potential assignment $W_{t}(w_{1:t-1})$ which would model the assignment $W_{t}$ that would be realized along the assignment path $w_{1:t-1} \in \cW^{t-1}$ and would open an \textit{indirect}, causal mechanism that allows the assignment $W_{t}$ to indirectly affect future outcomes through its effect on future assignments.\footnote{Such indirect causal mechanisms are often studied in a large biostatistics literature on longitudinal causal effects and dynamic treatment regimes -- e.g., see Chapter 19 of \cite{HernanRobins(18)}.}
The assignment process $\{W_t\}$ can still nonetheless have rich dependence. 
Assumption \ref{assumption: underlying processes} places no restrictions on how $W_{t}, W_{s}$ for $s \neq t$ are probabilistically related.

In focusing on the direct causal effects of assignments on outcomes, we adopt a common perspective in both macroeconometrics and financial econometrics. 
In particular, it is common in macroeconometrics to focus on studying the direct causal effects of underlying economic ``shocks'' on outcomes, which are thought to be underlying ``random causes'' that drive economic fluctuations and are causally unrelated to one another \citep{Frisch(33), Slutzky(37), Sims(80)}. 
The empirical goal is to therefore trace out the dynamic causal effects of these primitive, economic shocks $\{W_{t}\}$ on macroeconomic outcomes $\{Y_t\}$. 
We refer the readers to \cite{Ramey(16)}, \cite{StockWatson(16)}, and \cite{StockWatson(18)} for recent discussions of this perspective in macroeconometrics. 
We further discuss the connections between the assignments in a direct potential outcome system and economic ``shocks'' in Section \ref{sect:assignOut}.

\begin{remark}[Background processes]
    We could have further introduced the background process $\{X_t\}_{t \geq 1}$ that is causally unaffected by the assignment process. Such a process would play the same role as pre-assignment covariates in cross-sectional or longitudinal studies.
\end{remark}

\subsection{Dynamic Causal Effects}\label{sect:causaleffect}

Any comparison of the potential outcome process at a particular point in time along different possible realizations of the assignment process define a \textit{dynamic causal effect}. The dynamic causal effect at time $t$ for assignment path $w_{1:t} \in \mathcal{W}^{t}$ and counterfactual path $w_{1:t}^\prime \in \mathcal{W}^{t}$ is $Y_{t}(w_{1:t}) - Y_t(w_{1:t}^\prime)$.
Of course, this is an enormous class of dynamic causal effects as there are exponentially many possible paths $w_{1:t} \in \mathcal{W}^{t}$. We therefore introduce causal estimands that average over these dynamic causal effects along various underlying assignment paths.

To do so, let us introduce some shorthand. For $t \geq 1$, $h \geq 0$, and any fixed $w \in \mathcal{W}$, write the time-$(t+h)$ potential outcome at the assignment process $(W_{1:t-1}, w, W_{t+1:t+h})$ as $Y_{t+h}(w) := Y_{t+h}(W_{1:t-1}, w, W_{t+1:t+h})$.
Notice that $Y_{t+h} = Y_{t+h}(W_{t})$ in this notation.

\begin{definition}[Dynamic causal effects]\label{defn:impulse, filtered, average causal effect}
    For $t \geq 1$, $h \geq 0$, and any fixed $w, w^\prime \in \mathcal{W}$, the time-$t$, $h$-period ahead \textbf{impulse causal effect} is $Y_{t+h}(w) - Y_{t+h}(w^\prime)$, the \textbf{filtered treatment effect} is $\mathbb{E}[ Y_{t+h}(w) - Y_{t+h}(w^\prime) \mid \mathcal{F}_{t-1} ]$, and the \textbf{average treatment effect} is $\mathbb{E}[Y_{t+h}(w) - Y_{t+h}(w^\prime) ]$.
\end{definition}
\noindent The impulse causal effect measures the \textit{ceteris paribus} causal effect of intervening to switch the time-$t$ assignment from $w^\prime$ to $w$ on the $h$-period ahead outcomes holding all else fixed along the assignment process. The impulse causal effect is a random object since the potential outcome process, the past assignments $W_{1:t-1}$, and the future assignments $W_{t+1:t+h}$ are stochastic.
The \textit{filtered treatment effect} averages the impulse causal effect conditional on the natural filtration of assignments and observed outcomes up to time $t - 1$.\footnote{We use the nomenclature ``filtered'' following the stochastic process literature, where filtering refers to the sequential estimation of time-varying unobserved variables, e.g. Kalman filter \citep{Kalman(60), DurbinKoopman(12)}, particle filter \citep{GordonSalmondSmith(93), PittShephard(99jasa), ChopinPapasphiliopoulos(20)}, and hidden discrete Markov models \citep{BaumPetrie(66), Hamilton(89)}.}
Finally, the \textit{average treatment effect} further averages the filtered treatment effect over the filtration, yielding the unconditional expectation of the impulse causal effect $Y_{t+h}(w) - Y_{t+h}(w')$. 

We further define analogous versions of the dynamic causal effects for a particular scalar assignment. For any fixed $w_k \in \mathcal{W}_k$, define $Y_{t+h}(w_k) := Y_{t+h}(W_{1:t-1}, W_{1:k-1,t},w_k, W_{k+1:d_W, t}, W_{t+1:t+h})$.
The corresponding time-$t$, $h$-period ahead impulse causal effect, filtered treatment effect, and average treatment effect for the $k$-th assignment are, respectively: $Y_{t+h}(w_k) - Y_{t+h}(w_k^\prime)$, $\mathbb{E}[\{Y_{t+h}(w_k) - Y_{t+h}(w_k^\prime)\} \mid \mathcal{F}_{t-1}]$, and $\mathbb{E}[Y_{t+h}(w_k) - Y_{t+h}(w_k^\prime) ]$.
    
The dynamic causal effects in Definition \ref{defn:impulse, filtered, average causal effect} summarize the causal effect of discrete interventions to switch the time-$t$ assignments on the outcomes. We finally introduce derivatives that summarize marginal causal effects of incrementally varying the time-$t$ assignment.  

\begin{definition} If they individually exist, $Y'_{t+h}(w_k) = \frac{\partial Y_{t+h}(w_k)}{\partial w_k}$ is the time-$t$, $h$-period ahead \textbf{marginal impulse causal effect}, $\mathbb{E}[Y'_{t+h}(w_k)\mid \mathcal{F}_{t-1}]$ is the \textbf{marginal filtered treatment effect}, and $\mathbb{E}[Y'_{t+h}(w_k)]$ is the \textbf{marginal average treatment effect}.
\end{definition}

\subsection{Links to macroeconometrics}\label{section: links to macro}

Before continuing, we highlight how the direct potential outcome system naturally links to several recent developments and debates in macroeconometrics and encompasses many familiar parametric models in that field. 

First, the direct potential outcome system allows us to analyze what assumptions must be placed on the assignment process to endow causal meaning to common statistical estimands without resorting to functional form assumptions. 
Workhorse models in macroeconometrics, such as the structural vector moving average, assume linearity. 
However, this nullifies state-dependence and asymmetry in dynamic causal effects.
Researchers recognize the restrictiveness of linearity, yet attempt to weaken it on a case-by-case basis. 
For example, on the possible nonlinear effects of oil prices \citep{KillianVigfusson(11)-Oil, KillianVigfusson(11)-QE, Hamilton(11)-NonlinearityOil}; on the nonlinear and state dependent effects of monetary policy \citep{TenreyroThwaites(16), JordaSchularickTaylor(20), AruobaEtAl(21), Mavroeidis(21)}, and on state-dependent fiscal multipliers \citep{AuerbachGorodnichenk(12)-AEJ, AuerbachGorodnichenko(12)-NBER, RameyZubairy(18), CloyneEtAl(20)}.
Similarly, the direct potential outcome system does not rely on ``invertibility'' or ``recoverability'' assumptions about the assignment and potential outcome processes \citep{ChahrourJurado(21)}. Understanding what can be identified about dynamic causal effects without relying on these assumptions is an active area \citep{StockWatson(18), PlagborgMoller(19), PlagborgMollerWolf(20), ChahrourJurado(21), GoncalvesEtAl(21)}.

Second, a rapidly growing body of empirical research in macroeconometrics attempts to estimate dynamic causal efects in settings where researchers directly observe both the assignments and outcomes, $\{w_t^{obs},y_t^{obs}\}_{t \ge 1}$. 
In this line of work, empirical researchers creatively construct measures of the underlying economic shocks of interest $W_t$, and then use these constructed shocks to directly estimate dynamic causal effects on macroeconomic outcomes using reduced-form methods such as local projections \citep{Jorda(05)} or autoregressive distributed lag models \citep{BaekLee(21)}.
This line of work has recently been called ``direct causal inference'' by \cite{NakamuraSteinsson(18)-JEP}. 
The direct potential outcome system provides a causal foundation for such reduced-form methods in time series, elucidating the assumptions that the constructed shock must satisfy in order for the reduced form estimands to have a nonparametric causal interpretation.

\subsubsection{Examples from Macroeconomics}
Many leading causal models in macroeconomics can be cast as special cases of the direct potential outcome system by placing additional restrictions on the potential outcome process. 

\begin{example}[Structural vector moving average (SVMA) model]
The SVMA model is the leading workhorse model for studying dynamic causal effects in macroeconometrics \citep[e.g.,][]{KilianLutkepohl(17), StockWatson(18)}. Any infinite-order SVMA model can be expressed as a direct potential outcome system by assuming that the potential outcome process satisfies
\[
 Y_{t}(w_{1:t}) := \sum_{l=0}^{t-1} \Theta_{l} w_{t-l} + Y_t^*,
\]
where $\{W_t\}_{t \geq 1}$ is the assignment process, $\{\Theta_{l}\}_{0 \leq l < t}$ is a sequence of lag-coefficient matrices, and $\{Y_{t}^*\}_{t \geq 1}$ is a stochastic process that is causally unaffected by the assignment process. 
In this sense, the SVMA model imposes that the potential outcome process is linear in the assignment process.
This mapping requires no assumptions on the dimensionality of the assignment process $d_{w}$, the dimensionality of the potential outcome process $d_y$, nor the lag-coefficient matrices. 
As discussed in \cite{PlagborgMollerWolf(20)}, such an infinite-order SVMA model is consistent with all discrete-time Dynamic Stochastic General Equilibrium models as well as all stable, linear structural vector autoregression (SVAR) models. $\blacktriangle$
\end{example}

\begin{example}[Nonlinear structural vector autoregressions (SVAR)]
Recent advances in nonlinear SVARs can also be cast as special cases of the direct potential outcome system. 
As an illustration, consider the motivating example in \cite{GoncalvesEtAl(21)}:
    \begin{align*}
        Y_{1,t}(w_{1:t}) = w_{1,t},\quad Y_{2,t}(w_{1:t}) = b + \beta Y_{1,t}(w_{1:t}) + \rho Y_{2,t-1}(w_{1:t-1}) + c f(Y_{1,t}(w_{1:t})) + w_{2,t},
    \end{align*}
where $f(\cdot)$ is a nonlinear function. 
Given a stochastic initial condition $Y_{2,0} := \epsilon_{2,0}$ that is causally unaffected by the assignment process, iterating this system of equations forward arrives at a potential outcome process $Y_{1,t}(w_{1:t}) = w_{1,t},$ and $Y_{2,t}(w_{1:t}) = g_{2,t}(w_{1:t}, \epsilon_{2,0}; \theta)$, where $g_{2,t}(\cdot)$ is a known function and $\theta := (b, c, \beta, \rho)$ are the parameters.  
This is a direct potential outcome system where $Y_{1,t}(w_{1:t})$ is non-random and only depends on the contemporaneous assignment, and the randomness in $Y_{2:t}(w_{1:t})$ is driven by the initial condition. 
Other recent examples of nonlinear SVARs include \cite{AruobaEtAl(21)} and \cite{Mavroeidis(21)}. $\blacktriangle$
\end{example}

\begin{example}
\cite{AngristKuersteiner(11)} and \cite{AngristJordaKuersteiner(18)} introduce a potential outcome model for time series settings that is a special case of the direct potential outcome system. 
Using our notation, \cite{AngristKuersteiner(11)} introduce a system of structural equations in which for $t \geq 1$,
    \begin{align*}
        Y_{1,t}(w_{1:t}) = f_{1,t}(Y_{1:t-1}(w_{1:t-1}), w_{1,t}; \epsilon_{0}), \quad 
        Y_{2,t}(w_{1:t}) = f_{2,t}(Y_{1,t}(w_{1:t}), w_{2,t}, w_{1:t-1}; \epsilon_{0}),
    \end{align*}
where $f_{1,t}(\cdot), f_{2,t}(\cdot)$ are deterministic functions and $\epsilon_{0}$ is a random initial condition. 
These structural equations impose that $w_{1:t}$ only impacts $Y_{1,t}$ through $w_{1,t}$ directly and through $Y_{1,t-1}$ indirectly. Further, $w_{2,1:t}$ only impacts $Y_{2,t}$ contemporaneously. 
Related thinking includes \cite{WhiteKennedy(09)} and \cite{WhiteLu(10)}.  
Through forward iteration of the system, this can also be expressed as a direct potential outcome system. 
In this system of structural equations, the authors defined the collection of their time-$t+h$ potential outcomes, as $\{Y_{t+h}(w_{1:t-1}^{obs},w,W_{t+1:t+h}):w \in \mathcal{W}_W\}$ and focused on what they call the ``average policy effect'' $\mathbb{E}[Y_{t+h}(w_{1:t-1}^{obs},w,W_{t+1:t+h})-Y_{t+h}(w_{1:t-1}^{obs},w',W_{t+1:t+h})]$. $\blacktriangle$
\end{example}

\begin{example}[Expectations]
Macroeconomists often consider how assignments are influenced by the distribution of future outcomes and how they in turn vary with assignments. For example, consumers and firms are modelled as forward-looking and so, expectations about future outcomes influence behavior today. Consider a simple optimization-based version \citep[e.g.,][]{Lucas(72), Sargent(81)} in which the assignment process is given by  
    \begin{equation}\label{eqn:W by opt}
        W_t \in \underset{w_t}{\arg \max} \mbox{ }\underset{w_{t+1:T}}{\max} \mbox{ } \mathbb{E}[  \mbox{ } U(Y_{t:T}(w_{1:t-1}^{obs},w_{t:T}),w_{t:T}) \,|\, y_{1:t-1}^{obs},w_{1:t-1}^{obs}],
    \end{equation}
where $U$ is a utility function of future outcomes and assignments, while $\mathcal{F}_{t-1}$ is written out in long hand as $y_{1:t-1}^{obs},w_{1:t-1}^{obs}$. For each possible $w_{t:T}\in \mathcal{W}^{T-t+1}$, the expectation is over the law of $Y_{t:T}(w_{1:t-1}^{obs},w_{t:T})|y_{1:t-1}^{obs},w_{1:t-1}^{obs}.$ 
This decision rule delivers the output $\{W_t,Y_t(W_{1:t})\}_{t \geq 1}$. 
This looks like a direct potential outcome system since Assumption \ref{assumption: non-anticipating potential outcomes} holds.
%\footnote{The non-anticipation assumptions are similarly plausible if a different model for expectations is used. ``Natural expectations'' as in \cite{FusterLaibsonMendel(10)} or ``diagnostic expectations'' as in \cite{BordaloGennaioliShleifer(18)} both only allow current decisions to depend on (possibly biased) beliefs about future outcomes, not the exact realizations along alternative paths $Y_{t}(w_{1:t})$.} 
The assignment $W_{t}$ could be a deterministic function of past data if the optimal choice is unique, which would violate Assumption \ref{asm: sequentially probabilistic treatment}. However, incorporating noise in the decision rule (\ref{eqn:W by opt}) would deliver a direct potential outcome system. $\blacktriangle$
\end{example}

\section{Estimands Based on Assignments and Outcomes}\label{sect:assignOut}

In this section, we establish conditions under which common statistical estimands based on assignments and outcomes have causal meaning in the direct potential outcome system $\{W_t,\{Y_t(w_{1:t}):w_{1:t}\in \mathcal{W}^t\}\}_{t \ge 1}$, where researchers observe the realized assignments and realized outcomes $\{w_t^{obs},y_t^{obs}\}_{t\ge 1}$. 
We study the following statistical estimands: the impulse response function, local projections, the generalized impulse response function and the local filtered projection.
Table \ref{tab:assignmentoutcome} defines these estimands and summarizes our main results on their causal interpretation under important restrictions on the assignment process and other technical conditions. 
In this section, there is no loss in generality in assuming the outcome is scalar. The more general case is covered by running the analysis equation by equation.  

\begin{table}[htbp!]
{\footnotesize
   \centering
    \begin{tabular}{l || l|l}
    \textbf{Name}  & \textbf{Estimand} & \textbf{Causal Interpretation}  \\ \hline \hline
        Impulse Response& $\mathbb{E}[ Y_{t+h} \mid W_{k,t} = w_k] $ & 
        $\mathbb{E}[Y_{t+h}(w_k)- Y_{t+h}(w_k^\prime)];$ \\
        Function &$- \mathbb{E}[ Y_{t+h} \mid W_{k,t} = w_k^\prime]$&  
        \\
        &&\\
        Local Projection  &  $\frac{ Cov(Y_{t+h}, W_{k,t}) }{ Var(W_{k,t}) }$ & $\frac{\int_{\mathcal{W}_k} \mathbb{E}[Y'_{t+h}(w_{k})] \mathbb{E}[G_t(w_k)] dw_k}{\int_{\mathcal{W}_k} \mathbb{E}[G_t(w_k)]dw_k}$ \\
        &&\\
        Generalized Impulse
        & $\mathbb{E}[ Y_{t+h} \mid W_{k,t} = w_k, \mathcal{F}_{t-1}]$ &  $\mathbb{E}[Y_{t+h}(w_k)- Y_{t+h}(w_k^\prime) \mid \mathcal{F}_{t-1}];$ \\
        Response Function &$-\mathbb{E}[ Y_{t+h} \mid W_{k,t} = w_k^\prime, \mathcal{F}_{t-1}]$& \\
        && \\
%        Generalized Local &$\frac{ Cov(Y_{t+h}, W_{k,t} \mid \mathcal{F}_{t-1}) }{ Var(W_{k,t}\mid \mathcal{F}_{t-1}) }$& $\frac{\int_{\mathcal{W}_k} \mathbb{E}[Y'_{t+h}(w_{k}) \mid \mathcal{F}_{t-1}] \mathbb{E}[G_{t|t-1}(w_k)\mid \mathcal{F}_{t-1}] dw_k}{\int_{\mathcal{W}_k} \mathbb{E}[G_{t|t-1}(w_k)\mid \mathcal{F}_{t-1}]dw_k}$\\
%        Projection && \\
%        && \\
        Local Filtered Projection &$\frac{ \mathbb{E}[\{Y_{t+h}-\hat{Y}_{t+h|t-1}\}\{W_{k,t} -\hat{W}_{k,t|t-1}\}] }{ \mathbb{E}[\{W_{k,t} -\hat{W}_{k,t|t-1}\}^2] }$& $\frac{\int_{\mathcal{W}_k} \mathbb{E} \left[ \mathbb{E}[Y'_{t+h}(w_{k}) \mid \mathcal{F}_{t-1}] \mathbb{E}[G_{t|t-1}(w_k)\mid \mathcal{F}_{t-1}] \right] dw_k}{\int_{\mathcal{W}_k} \mathbb{E}[G_{t|t-1}(w_k)]dw_k}$
    \end{tabular}
    \caption{Top line results for the causal interpretation of common estimands based on assignments and outcomes.  Here $h \geq 0$, $w_k, w_k^\prime \in \mathcal{W}_k$, $G_t(w_k)= 1\{w_k \leq W_{k,t}\} (W_{k,t} - \mathbb{E}[W_{k,t}])$ and $G_{t|t-1}(w_k)= 1\{w_k \leq W_{k,t}\} (W_{k,t} - \mathbb{E}[W_{k,t}\mid \mathcal{F}_{t-1}])$, while $\hat{Y}_{t+h|t-1} := \mathbb{E}[Y_{t+h}\mid \mathcal{F}_{t-1}]$ and  $\hat{W}_{k, t\mid t - 1} := \mathbb{E}[W_{k,t} \mid \mathcal{F}_{t-1}]$.  Note that $\mathbb{E}[G_t(w_k)] \ge 0$ and $\mathbb{E}[G_{t|t-1}(w_k)\mid \mathcal{F}_{t-1}] \ge 0$.}
   \label{tab:assignmentoutcome}
    }
\end{table}

\subsection{Impulse Response Function}

We first study conditions under which the unconditional \textit{impulse response function} \citep{Sims(80)} is the $h$-period ahead average treatment effect. For $h \geq 0$ and deterministic $w_k, w_k^\prime \in \mathcal{W}_{k}$, the impulse response function is defined by, if it exists,
    \begin{equation}
        IRF_{k,t,h}(w_k, w_k^\prime) := \mathbb{E}[ Y_{t+h} \mid W_{k,t} = w_k] - \mathbb{E}[ Y_{t+h} \mid W_{k,t} = w_k^\prime].
    \end{equation}
$IRF_{k,t,h}(w_k, w_k^\prime)$ can be decomposed into the average treatment effect and a selection bias term.

\begin{theorem}\label{theorem: IRF decomposition under full obs}
    Assume a direct potential outcome system, consider some $k = 1, \hdots, d_w$, $t \geq 1$, $h \geq 0$, fix $w_k, w_k^\prime \in \mathcal{W}_k$ and that $\mathbb{E}[ | Y_{t+h}(w_k) - Y_{t+h}(w_k')| ] < \infty$. Then, 
        \[
        IRF_{k,t,h}(w_k, w_k^\prime) = \mathbb{E}[Y_{t+h}(w_k)- Y_{t+h}(w_k^\prime)] + \Delta_{k,t,h}(w_k, w_k^\prime),
        \]
    where 
        \[
        \Delta_{k,t,h}(w_k, w_k^\prime) := \frac{ Cov\left( Y_{t+h}(w_k), 1\{W_{k,t} = w_k\} \right)}{ \mathbb{E}[ 1\{W_{k,t} = w_k\}] } - \frac{ Cov\left( Y_{t+h}(w_k'), 1\{W_{k,t} = w_k^\prime\} \right)}{ \mathbb{E}[ 1\{W_{k,t} = w_k^\prime\}] }.
        \]    
\end{theorem}

\noindent The impulse response function therefore equals the average treatment effect if and only if the selection bias term $\Delta_{k,t,h}(w_k, w_k^\prime)=0.$ A sufficient condition for this to hold is that the two covariance terms are zero. 

Notice that these covariance terms depend on how the assignment $W_{k,t}$ covaries with the potential outcome $Y_{t+h}(w_k)$. Since $Y_{t+h}(w_k) := Y_{t+h}(W_{1:t-1}, W_{1:k-1,t},w_{k}, W_{k+1:d_W,t}, W_{t+1:t+h})$ by definition, the selection bias therefore depends on how the assignment $W_{k,t}$ relates to (i) past assignments $W_{1:t-1}$, (ii) other contemporaneous assignments $W_{1:k-1,t},W_{k+1:d_W,t}$, (iii) future assignments $W_{t+1:t+h}$, and (iv) the potential outcome process $Y_{t+h}(w_{1:t+h})$.
By placing restrictions on the assignment process, we arrive at sufficient conditions for $\Delta_{k,t,h}(w_k, w_k^\prime)$ to be zero.  
\begin{theorem}\label{theorem: shocks, IRFs}
Under the same conditions as Theorem \ref{theorem: IRF decomposition under full obs}, if 
\begin{equation}\label{eqn:cov2}
    Cov\left( Y_{t+h}(w_k), 1\{W_{k,t} = w_k\} \right)=0, \mbox{ and } Cov\left( Y_{t+h}(w_k'), 1\{W_{k,t} = w_k'\} \right)=0
\end{equation}
then $\Delta_{k,t,h}(w_k, w_k^\prime) = 0$. Moreover, Equation (\ref{eqn:cov2}) is satisfied if 
\begin{equation}\label{eqn:condindep0}
    W_{k,t} \indep \left( W_{1:t-1}, W_{1:k-1,t},W_{k+1:d_W,t}, W_{t+1:t+h}, \{ Y_{t+h}(w_{1:t+h}) \colon w_{1:t+h} \in \mathcal{W}^{t+h} \} \right).
\end{equation}
\end{theorem} 

\noindent Equation (\ref{eqn:condindep0}) says the selection bias is zero if the assignment $W_{k,t}$ is randomized in the sense that it is independent of all other assignments and the time-$(t+h)$ potential outcomes.  

Recent reviews on dynamic causal effects in macroeconometrics by \cite{Ramey(16)} and \cite{StockWatson(18)} argue intuitively that the impulse response function of observed outcomes to ``shocks'' in parametric structural models, such as the SVMA, are analogous to an average treatment effect in a randomized experiment from cross-sectional causal inference.\footnote{\cite{StockWatson(18)} write on pg. 922: ``The macroeconometric jargon for this random treatment is a 'structural shock:' a primitive, unanticipated economic force, or driving impulse, that is unforecastable and uncorrelated with other shocks. The macroeconomist's shock is the microeconomists' random treatment, and impulse response functions are the causal effects of those treatments on variables of interest over time, that is, dynamic causal effects.''} However, these statements rely on either intuitive descriptions of the statistical properties of shocks\footnote{\cite{Ramey(16)} writes on pg. 75, ``the shocks should have the following characteristics: (1) they should be exogenous with respect to the other current and lagged endogenous variables in the model; (2) they should be uncorrelated with other exogenous shocks; otherwise, we cannot identify the unique causal effects of one exogenous shock relative to another; and (3) they should represent either unanticipated movements in exogenous variables or news about future movements in exogenous variables.''}, or on a specific parametric model for the potential outcome process to link the impulse response function to an average dynamic causal effect. 
Theorem \ref{theorem: shocks, IRFs} clarifies that if the assignment $W_{k,t}$ is randomly assigned in these sense of Equation (\ref{eqn:condindep0}), then the impulse response function identifies an average treatment effect in the direct potential outcome system.
In this sense, Theorem \ref{theorem: shocks, IRFs} provides an interpretation of ``shock'' in terms of a random assignment assumption in a direct potential outcome system.

Furthermore, Theorems \ref{theorem: IRF decomposition under full obs}-\ref{theorem: shocks, IRFs} clarifies a recent empirical literature that seeks to directly construct measures of the shocks of interest and measure dynamic causal effects through reduced-form estimates of impulse response functions --- so called ``direct causal inference'' \citep[e.g., see][]{NakamuraSteinsson(18)-JEP, BaekLee(21)}. 
In order for researchers to causally interpret reduced-form impulse response functions of outcomes on particular constructed shocks as nonparametrically identifying an average treatment effect, then the constructed shocks must be randomized in these sense given in Theorem \ref{theorem: shocks, IRFs}.

\subsection{Local Projection Estimand}
Under the conditions of Theorem \ref{theorem: IRF decomposition under full obs}, impulse response functions are causal, but nonparametrically estimating impulse response functions is in general challenging. If the assignment is observed by the researcher, it is therefore common to estimate impulse response functions using ``local projections'' \citep{Jorda(05)}, which directly regresses the $h$-step ahead outcome on a constant and the assignment.
The corresponding local projection estimand is
    \begin{equation}
        LP_{k,t,h} := \frac{ Cov(Y_{t+h}, W_{k,t}) }{ Var(W_{k,t}) }.
    \end{equation}
Theorem \ref{theorem: local projection estimand, IRF} establishes that $LP_{k,t,h}$ identifies a weighted average of marginal causal effects of the assignment on the $h$-step ahead outcome.

\begin{theorem}\label{theorem: local projection estimand, IRF}
    Under the same conditions as Theorem \ref{theorem: IRF decomposition under full obs}, further assume that: 
        \begin{enumerate}
            \item[i.] The support of $W_{k,t}$ is a closed interval, $\mathcal{W}_{k} := [\underline{w}_k, \overline{w}_k] \subset \mathbb{R}$.  
            
            \item[ii.] \underline{Differentiability}: $Y_{t+h}(w_{k})$ is continuously differentiable in $w_k$, as is $\mathbb{E}[Y'_{t+h}(w_{k})]$.
            
            \item[iii.] \underline{Independence}: $W_{k,t} \indep \{Y_{t+h}(w_k) \colon w_k \in \mathcal{W}_k\}$.   
            
        \end{enumerate}
    Then, if it exists,
    $$ LP_{k,t,h} = \frac{\int_{\mathcal{W}_k} \mathbb{E}[Y'_{t+h}(w_{k})] \mathbb{E}[G_t(w_k)] dw_k}{\int_{\mathcal{W}_k} \mathbb{E}[G_t(w_k)]dw_k},$$
    where $G_t(w_k)= 1\{w_k \leq W_{k,t}\} (W_{k,t} - \mathbb{E}[W_{k,t}])$, noting $\mathbb{E}[G_t(w_k)] \ge 0$.
\end{theorem}

\noindent The local projection estimand $LP_{k,t,h}$ is therefore a weighted average of marginal average treatment effects of $W_{k,t}$ on the $Y_{t+h}$, where 
the weights $\mathbb{E}[G_t(w_k)]$ are non-negative and sum to one. Thus, if the assignment $W_{k,t}$ is a shock in the sense stated in Theorem \ref{theorem: shocks, IRFs}, the local projection estimand also has a nonparametric causal interpretation. 
Recently, \citet[][]{KolesarPlagborgMoller(24)} build on this identification result to show that researchers can estimate the implied weights and report them as a valuable causal diagnostic in empirical work.

\subsection{Generalized Impulse Response Function}
In non-linear time series models, it is common to focus on the conditional version of the impulse response function, the $h$-period ahead \textit{generalized impulse response function} \citep{GallantRossiTauchen(93), KoopPesaranPotter(96),GourierouxJasiak(05)}, which is
    \begin{equation}
        GIRF_{k,t,h}(w_k, w_k^\prime \mid \mathcal{F}_{t-1}) := \mathbb{E}[ Y_{t+h} \mid W_{k,t} = w_k, \mathcal{F}_{t-1}] - \mathbb{E}[ Y_{t+h} \mid W_{k,t} = w_k^\prime, \mathcal{F}_{t-1}].
    \end{equation}
Mirroring our analysis of the impulse response function, we next show that $GIRF_{k,t,h}$ can be decomposed into the filtered treatment effect and a selection bias term.

\begin{theorem}\label{theorem: GIRF decomposition under full obs}
    Assume a direct potential outcome system, some $k = 1, \hdots, d_w$, $t \geq 1$, and $h \geq 0$ and that $\mathbb{E}[ |Y_{t+h}(w_k) - Y_{t+h}(w_k^\prime)| \mid \mathcal{F}_{t-1}] < \infty$. Then, for any deterministic $w_k, w_k^\prime \in \mathcal{W}$,
        \[
        GIRF_{k,t,h}(w_k, w_k^\prime \mid \mathcal{F}_{t-1}) = \mathbb{E}[\{Y_{t+h}(w_k) - Y_{t+h}(w_k^\prime)\} \mid \mathcal{F}_{t-1} ] + \Delta_{k,t,h}(w_k, w_k^\prime \mid \mathcal{F}_{t-1}),
        \]
    where 
        \[ \Delta_{k,t,h}(w_k, w_k^\prime \mid \mathcal{F}_{t-1}) := \frac{ Cov\left( Y_{t+h}(w_k), 1\{W_{k,t} = w_k\} \mid \mathcal{F}_{t-1} \right)}{ \mathbb{E}[ 1\{W_{k,t} = w_k\} \mid \mathcal{F}_{t-1}] } - \frac{ Cov\left( Y_{t+h}(w_k^\prime), 1\{W_{k,t} = w_k^\prime\} \mid \mathcal{F}_{t-1} \right)}{ \mathbb{E}[ 1\{W_{k,t} = w_k^\prime\} \mid \mathcal{F}_{t-1}] }.
        \]
\end{theorem}

% Sufficient conditions for the selection bias term $\Delta_{k,t,h}(w_k, w_k^\prime \mid \mathcal{F}_{t-1})$ to equal zero is that the two conditional covariances are zero.  
% Repeating the unconditional case, Theorem \ref{theorem: shocks, GIRFs} provides sufficient conditions such that the selection bias term is equal to zero. 

\begin{theorem}\label{theorem: shocks, GIRFs}
Under the same conditions as Theorem \ref{theorem: GIRF decomposition under full obs}, if 
\begin{equation}\label{eqn:cov3} 
Cov\left( Y_{t+h}(w_k), 1\{W_{k,t} = w_k\} \mid \mathcal{F}_{t-1}\right)=0,\quad Cov\left( Y_{t+h}(w_k'), 1\{W_{k,t} = w_k'\} \mid \mathcal{F}_{t-1} \right)=0, 
\end{equation}
then $\Delta_{k,t,h}(w_k, w_k^\prime) = 0$. 
Moreover, Equation (\ref{eqn:cov3}) is implied by 
 \begin{equation}\label{eqn:blockseqrandom}
        [W_{k,t} \indep \{Y_{t+h}(w_{k}) \colon w_{k} \in \mathcal{W}_k\}] \mid \mathcal{F}_{t-1},
    \end{equation}
    which is in turn implied by 
        \begin{equation}\label{eqn:condindep2}
            [W_{k,t} \indep \left(W_{1:k-1,t},W_{k+1:d_W,t}, W_{t+1:t+h}, \{Y_{t+h}(w_{1:t-1}^{obs}, w_{t:t+h}) \colon w_{t:t+h} \in \mathcal{W}^{h+1}\} \right)] \mid \mathcal{F}_{t-1}.
        \end{equation}
\end{theorem}

\noindent Therefore, under Equation (\ref{eqn:cov3}), the selection bias $\Delta_{k,t,h}(w_k, w_k^\prime \mid \mathcal{F}_{t-1}) = 0$ and the generalized impulse response function identifies the filtered impulse causal effect. 
At first glance, Equation (\ref{eqn:blockseqrandom}) appears analogous to a typical unconfoundedness assumption from cross-sectional causal inference or sequential randomization assumption from longitudinal causal inference. That is, it imposes that conditional on the history up to time $t- 1$, the assignment $W_{k,t}$ must be as good as randomly assigned. However, recall that the notation $Y_{t+h}(w_k)$ buries dependence on (i) other contemporaneous assignments $W_{1:k-1,t},W_{k+1:d_W,t}$; (ii) future assignments $W_{t+1:t+h}$; and (iii) the potential outcomes at time-$(t+h)$. Therefore, Equation (\ref{eqn:condindep2}) in Theorem \ref{theorem: shocks, GIRFs} provides further sufficient conditions, highlighting that it is sufficient to further impose that the assignment $W_{k,t}$ is jointly independent of all other contemporaneous and future assignments as well as the underlying potential outcomes. 
Finally, notice how much weaker Equation (\ref{eqn:condindep2}) is than Equation (\ref{eqn:condindep0}) as it allows the assignment to probabilistically depend flexibly on the past realized potential outcomes and realized assignments. 

\begin{remark}
How do the conditions in Theorem \ref{theorem: shocks, IRFs} relate to the conditions in Theorem \ref{theorem: shocks, GIRFs}? 
Applying the law of total covariance yields
\begin{align*}
Cov(Y_{t+h}(w_k),1\{W_{k,t} = w_k\}) 
&= \mathbb{E}[Cov(Y_{t+h}(w_k),1\{W_{k,t} = w_k\}\mid \mathcal{F}_{t-1})] \\
&+ Cov(\mathbb{E}[Y_{t+h}(w_k)\mid \mathcal{F}_{t-1}],\mathbb{E}[1\{W_{k,t} = w_k\}\mid \mathcal{F}_{t-1}]),
\end{align*}
so $Cov(Y_{t+h}(w_k),1\{W_{k,t} = w_k\})=0$ neither implies or is implied by  $Cov(Y_{t+h}(w_k),1\{W_{k,t} = w_k\}\mid \mathcal{F}_{t-1})]=0$. Hence, the conditional and unconditional cases are non-nested. 
If we instead work probabilistically, then the condition
    \[
    W_{k,t} \indep \left( W_{1:t-1}, W_{1:k-1,t},W_{k+1:d_W,t}, W_{t+1:t+h}, \{ Y_{1:t+h}(w_{1:t+h}) \colon w_{1:t+h} \in \mathcal{W}^{t+h} \} \right),
    \]
which strengthens Equation (\ref{eqn:condindep0}) to additionally require independence of the full potential outcome process, implies the Equation (\ref{eqn:condindep2}).
This second point is important practically. 
The generalized impulse response function identifies the filtered treatment effect provided that $[W_{k,t} \indep \{Y_{t+h}(w_{k}) \colon w_{k} \in \mathcal{W}_k\}]
\mid \mathcal{F}_{t-1}$. 
A temporally averaged generalized impulse response function recovers the average treatment effect without the need to employ the harsher condition $[W_{k,t} \indep \{Y_{t+h}(w_{k}) \colon w_{k} \in \mathcal{W}_k\}]$.
\end{remark}

\subsection{Generalized Local Projection and Local Filtered Projection Estimands}
Again estimating generalized impulse response functions nonparametrically is challenging. 
Under the same conditions as Theorem \ref{theorem: local projection estimand, IRF} but replacing condition (iii) with Equation (\ref{eqn:blockseqrandom}), the \textit{generalized local projection} satisfies
$$
\frac{ Cov(Y_{t+h}, W_{k,t} \mid \mathcal{F}_{t-1} ) }{ Var(W_{k,t} \mid \mathcal{F}_{t-1}) } = \frac{\int_{\mathcal{W}_k} \mathbb{E}[Y'_{t+h}(w_{k}) \mid \mathcal{F}_{t-1}] \mathbb{E}[G_{t|t-1}(w_k)\mid \mathcal{F}_{t-1}] dw_k}{\int_{\mathcal{W}_k} \mathbb{E}[G_{t|t-1}(w_k)\mid \mathcal{F}_{t-1}]dw_k},$$
where $G_{t|t-1}(w_k)= 1\{w_k \leq W_{k,t}\} (W_{k,t} - \mathbb{E}[W_{k,t}\mid \mathcal{F}_{t-1}])$, noting $\mathbb{E}[G_{t|t-1}(w_k)\mid \mathcal{F}_{t-1}] \ge 0$.
The generalized local projection is equivalent to a weighted average of conditional average marginal effects of $W_{k,t}$ on $Y_{t+h}$, where the weights now depend on the natural filtration but still are non-negative and sum to one.
    
Of more practical importance is the local projection of $Y_{t+h}-\hat{Y}_{t+h|t-1}$ on $W_{k,t}-\hat{W}_{k,t|t-1}$, where $\hat{Y}_{t+h \mid t-1} := \mathbb{E}[Y_{t+h} \mid \mathcal{F}_{t-1}]$ and $\hat{W}_{k, t\mid t - 1} := \mathbb{E}[W_{k,t} \mid \mathcal{F}_{t-1}].$  We call the associated estimand the \textit{local filtered projection}, which is defined as
    \begin{align*}
    \frac{ \mathbb{E}[\{Y_{t+h}-\hat{Y}_{t+h|t-1}\}\{W_{k,t} -\hat{W}_{k,t|t-1}\}] }{ \mathbb{E}[\{W_{k,t} -\hat{W}_{k,t|t-1}\}^2] }. 
    \end{align*}
Under the same conditions as needed for the generalized local projection plus needing the unconditional expectations to exist, the local filtered projection estimand equals
    \begin{align*}
    \frac{\int_{\mathcal{W}_k} \mathbb{E} \left[ \mathbb{E}[Y'_{t+h}(w_{k}) \mid \mathcal{F}_{t-1}] \mathbb{E}[G_{t|t-1}(w_k)\mid \mathcal{F}_{t-1}] \right] dw_k}{\int_{\mathcal{W}_k} \mathbb{E}[G_{t|t-1}(w_k)]dw_k}.
    \end{align*}
%where $G_{t|t-1}(w_k)= 1\{w_k \leq W_{k,t}\} (W_{k,t} - \mathbb{E}[W_{k,t}\mid \mathcal{F}_{t-1}])$.  
This is a long-run weighted average of the marginal filtered causal effect. The weights are non-zero and average to one over time. 
 
\section{The Instrumented Potential Outcome System}\label{sect:instruPOS}
We now use a special case of the direct potential outcome system to incorporate instrumental variables for the assignment process.
This is useful as a rapidly growing literature in macroeconomics exploits the use of instruments to identify dynamic causal effects \citep[e.g., see][among many others]{JordaSchularickTaylor(15), GertlerKaradi(15), RameyZubairy(18), StockWatson(18), PlagborgMollerWolf(20), JordaSchularickTaylor(20)}. 
Section \ref{sect:alldata} details the case where the researcher observes the assignments, the instruments, and the outcomes. 
Section \ref{sect:instrOutcome} considers the case where only the instruments and the outcomes are observed. 
    
\subsection{The Instrumented System}
We start by setting up an ``augmented assignment'' $V_t$, so that  $\left\{ V_{t},\left\{ Y_{t}(v_{1:t}):v_{1:t}\in \mathcal{W}_{V}^{t}\right\} \right\} _{t\geq 1}$ is a direct potential outcome system. 
The instrumented potential outcome system then imposes two further assumptions on the potential outcome system:  (i) that $\{V_t\}_{t \ge 1}$ splits into an ``instrument'' $\{Z_{t}\}_{t \ge 1}$ and a ``potential assignment'' $\{W_{t}(z_{t}): z_{t} \in \mathcal{W}_Z^t\}_{t \ge 1}$ that is only causally affected by the contemporaneous instrument, meaning $V_t=(Z_t,W_t(Z_t))$; (ii) the potential outcome process is only affected by the assignment $W_{1:t}$. 

\begin{definition}[Instrumented potential outcome system]
\label{Ex:IV}Assume $W_{t}\in \mathcal{W}_{W}$, $Z_{t}\in \mathcal{W}_{Z}$ and write $V_{t}=(W_{t},Z_{t})$. Assume 
$
\left\{ V_{t}, \left\{ Y_{t}(v_{1:t}):v_{1:t}\in \mathcal{W}_{W}^{t} \times \mathcal{W}_{Z}^{t} \right\} \right\} _{t\geq 1}
$
is a direct potential outcome system, and we additionally enforce the following assumptions:
\begin{enumerate}
\item[i.] \textbf{Contemporaneous Instrument}: The ``potential assignments'' satisfy
    \begin{align*}
    & W_{k,t}(\left\{z_{s}\right\}_{s\geq 1}) = W_{k,t}(z_{1:t-1}',z_t,\left\{ z_{s}^{\prime }\right\} _{s\geq t+1}) \\
    & W_{1:k-1,t}(\{z_{s}\}_{s\geq 1}) = W_{1:k-1,t}(\{z_{s}'\}_{s\geq 1}) \\
    & W_{k+1:d_W,t}(\{z_{s}\}_{s\geq 1}) = W_{k+1:d_W,t}(\{z_{s}'\}_{s\geq 1})
    \end{align*}
almost surely, for all $t \geq 1$ and all deterministic $\left\{ z_{t}\right\} _{t\geq 1}$ and $\left\{ z_{t}^{\prime }\right\} _{t\geq 1}$. Write the potential assignments as $\{W_{t}(z_t)=(W_{1:k-1,t},W_{k,t}(z_t),W_{k+1:d_W,t}):z_t\in \mathcal{W}_Z\}$,
while the assignment is $W_{t}= W_t(Z_t) = (W_{1:k-1,t},W_{k,t}(Z_t),W_{k+1:d_W,t}).$

\item[ii.] \textbf{Potential Outcome Exclusion}: $Y_{t}(\left( w_{1},z_{1}\right) ,...,\left( w_{t},z_{t}\right))=Y_{t}(\left( w_{1},z_{1}^{\prime }\right) ,...,\left( w_{t},z_{t}^{\prime}\right) )$ almost surely for all $w_{1:t}\in \mathcal{W}_{W}^{t}$ and $z_{1:t},z_{1:t}^{\prime } \in \mathcal{W}_{Z}^{t}$. 
Write the potential outcomes as $\left\{Y_{t}(w_{1:t}):w_{1:t}\in \mathcal{W}_{W}^{t}\right\} $ and outcome as $Y_{t}=Y_{t}(W_{1:t})$.

\item[iii.] \textbf{Output}: The output is $\{Z_t,W_t,Y_t\}_{t \ge 1} = \{Z_t,W_t(Z_t),Y_t(W_{1:t})\}_{t\ge1},$ while $Z_t$ and $\{Z_t\}_{t \ge 1}$ are called the ``contemporaneous instrument,'' and instrument process, respectively. 
\end{enumerate}

\noindent Any $\{Z_t,\{W_t(z_t),z_t\in \mathcal{W}_Z\},\{Y_t(w_{1:t}),w_{1:t}\in \mathcal{W}_W^t\}\}_{t \ge 1}$ satisfying (i)-(iii) is an {\bf instrumented potential outcome system}.
\end{definition}
% \noindent The simplest case is when both the assignment and instrument are scalar and binary, $\mathcal{W}_{W} = \{0, 1\}, \mathcal{W}_{Z} = \{0, 1\}$. In this case, the instrument $Z_t=1$  corresponds to ``intention to treat'' and $Z_t=0$ is ``intention to control.'' There is treatment and control as intended when $W_t(1)=1$ and $W_{t}(0) = 0$. But there can be noncompliance when $W_t(1)=0$ and $W_t(0)=1$. 

Assumption (i) imposes that $Z_{t}$ is only an instrument for the time-$t$, $k$-th assignment. This formalizes common empirical intuitions in macroeconometrics where a constructed external instrument is often ``targeted'' towards a single economic shock of interest -- for example, empirical researchers  construct proxies for a monetary policy shock \citep[e.g.,][]{GertlerKaradi(15), NakamuraSteinsson(18)-QJE, JordaSchularickTaylor(20)} or a fiscal policy shock \citep{RameyZubairy(18)}.
Assumption (ii) is the familiar outcome exclusion restriction on the instrument.

% To use this structure, we also need a type of ``relevance'' condition on the instrument. Such conditions will be stated as needed below.  

\section{Estimands Based on Assignments, Instruments and Outcomes}\label{sect:alldata}

We now study the conditions under which leading statistical estimands based on assignments, instruments and outcomes have causal meaning in the context of an instrumented potential outcome system $\{Z_t,\{W_t(z_t),z_t\in \mathcal{W}_Z\},\{Y_t(w_{1:t}),w_{1:t}\in \mathcal{W}_W^t\}\}_{t \ge 1}$. We consider the case in which the researcher observes the instruments, the assignments, and the outcomes $\{z_t^{obs},w_t^{obs},y_t^{obs}\}_{t\ge 1}.$

Since the assignments themselves are assumed to be directly observable, we focus on dynamic IV estimands that involve taking the ratio of an impulse response function of the outcome on the instrument relative to the impulse response function of the assignment on the instrument. 
We show that such dynamic IV estimands identify local average impulse causal effects in the sense of \cite{ImbensAngrist(94)}, \cite{AngristImbensRubin(96)}, and \cite{AngristGraddyImbens(00)-cts_late}. 
Our results in this section are most closely related to \cite{JordaSchularickTaylor(20)}, who used a potential outcome model analogous to that introduced in \cite{AngristJordaKuersteiner(18)}, to understand the causal content of local projection IV with a binary assignment and binary instrument. 
Specifically, we study the following statistical estimands: the Wald estimand, the IV estimand, the generalized Wald estimand, and the filtered IV estimand.
Table \ref{tab:assignmentinstrumentsoutcome} defines these estimands and summarizes our main results on their causal interpretation under important restrictions on the assignment process and other technical conditions.

\begin{table}[htbp!]
%{\footnotesize
   \centering
    \begin{tabular}{l || l|l}
    \textbf{Name}  & \textbf{Estimand} & \textbf{Causal Interpretation} \\ \hline \hline
         Wald&$\frac{ \mathbb{E}[Y_{t+h} \mid Z_{t} = z] - \mathbb{E}[ Y_{t+h} \mid Z_{t} = z^\prime ] }{ \mathbb{E}[W_{k,t} \mid Z_{t} = z] - \mathbb{E}[ W_{k, t} \mid Z_{t} = z^\prime] }$ & $\frac{\int_{\cW}   \mathbb{E}[Y'_{t+h}(w_{k})|H_t(w_k)=1]   \mathbb{E}[H_t(w_k)] dw_k}
        {\int_{\cW} \mathbb{E}[H_t(w_k)] dw_k}$ \\
        && \\
        IV  & $\frac{Cov(Y_{t+h},Z_t)}{Cov(W_t,Z_t)}$ & $\frac{\int_{\mathcal{W}_Z} \mathbb{E}[Y_{t+h}'(z_t)] \mathbb{E}[G_t(z_t)] d z_t}{\int_{\mathcal{W}_Z} \mathbb{E}[W_{t}'(z_t)] \mathbb{E}[G_t(z_t)] d z_t}$ \\
        &&\\
        Generalized
        &$\frac{ \mathbb{E}[Y_{t+h} | Z_{t} = z, \mathcal{F}_{t-1}] - \mathbb{E}[ Y_{t+h} | Z_{t} = z^\prime, \mathcal{F}_{t-1} ] }{ \mathbb{E}[W_{k,t} | Z_{t} = z, \mathcal{F}_{t-1}] - \mathbb{E}[ W_{k, t} | Z_{t} = z^\prime, \mathcal{F}_{t-1}] }$
        &$\frac{\int_{\cW}   \mathbb{E}[Y'_{t+h}(w_{k}) | H_t(w_k)=1,\mathcal{F}_{t-1}]   \mathbb{E}[H_t(w_k)| \mathcal{F}_{t-1}] dw_k}
        {\int_{\cW} \mathbb{E}[H_t(w_k) | \mathcal{F}_{t-1}] dw_k}$ \\
        Wald && \\
        && \\
%        Generalized&$\frac{Cov(Y_{t+h},Z_t| \mathcal{F}_{t-1})}{Cov(W_t,Z_t| \mathcal{F}_{t-1})}$&$\frac{\int_{\mathcal{W}_Z} \mathbb{E}[Y_{*,t+h}'(z_t)| \mathcal{F}_{t-1}] \mathbb{E}[G_t(z_t)| \mathcal{F}_{t-1}] d z_t}{\int_{\mathcal{W}_Z} \mathbb{E}[W_{t}'(z_t)| \mathcal{F}_{t-1}] \mathbb{E}[G_t(z_t)| \mathcal{F}_{t-1}] d z_t}$ \\
%        IV &&  \\
%        && \\
        Filtered IV &$\frac{\mathbb{E}[(Y_{t+h} - \hat{Y}_{t+h \mid t - 1})(Z_{t} - \hat{Z}_{t \mid t - 1})]}{\mathbb{E}[(W_{k,t} - \hat{W}_{k,t \mid t-1})(Z_{t} - \hat{Z}_{t \mid t-1})]}$& $\frac{\int_{\mathcal{W}_Z} \mathbb{E}[\mathbb{E}[Y_{t+h}'(z_t)| \mathcal{F}_{t-1}] \mathbb{E}[G_t(z_t)| \mathcal{F}_{t-1}]] d z_t}{\int_{\mathcal{W}_Z} \mathbb{E}[\mathbb{E}[W_{t}'(z_t)| \mathcal{F}_{t-1}] \mathbb{E}[G_t(z_t)| \mathcal{F}_{t-1}]] d z_t}$ \\
    \end{tabular}
    \caption{Top line results for the causal interpretation of common estimands based on assignments, instruments and outcomes.  Here $h \geq 0$, $z, z^\prime \in \mathcal{W}_{Z}$, $Y_{t+h}(z_t) := Y_{t+h}(W_{1:t-1},W_{t,1:k-1},W_k(z_t),W_{t,k+1:d_W},W_{t+1:t+h})$, $Y_{t+h}'(z_t) := \partial Y_{t+h}(z_t)/\partial z_t$, $H_t(w_k) = 1\{W_{k,t}(z^\prime) \leq w_k \leq W_{k,t}(z)\}$, $G_t(z_t) = 1\{z_t \le Z_t\}(Z_t-\mathbb{E}[Z_t])$ and $G_{t|t-1}(z_t) = 1\{z_t \le Z_t\}(Z_t-\mathbb{E}[Z_t] \mid \mathcal{F}_{t-1}])$, while $\hat{Y}_{t+h|t-1} = \mathbb{E}[Y_{t+h}\mid \mathcal{F}_{t-1}]$, $\hat{Z}_{t \mid t - 1} = \mathbb{E}[Z_t\mid \mathcal{F}_{t-1}]$ and $\hat{W}_{k, t \mid t-1} = \mathbb{E}[W_{k,t} \mid \mathcal{F}_{t-1}]$.  Note that $\mathbb{E}[G_t(z_t)] \ge 0$ and $\mathbb{E}[G_{t|t-1}(z_t)\mid \mathcal{F}_{t-1}] \ge 0$.}
   \label{tab:assignmentinstrumentsoutcome}
%    }
\end{table}

\subsection{Wald Estimand}\label{sect: unconditional IV estimand}

Consider the Wald estimand
    \[
    \frac{ \mathbb{E}[Y_{t+h} \mid Z_{t} = z] - \mathbb{E}[ Y_{t+h} \mid Z_{t} = z^\prime ] }{ \mathbb{E}[W_{k,t} \mid Z_{t} = z] - \mathbb{E}[ W_{k, t} \mid Z_{t} = z^\prime] }.
    \]
The numerator is the impulse response of the outcome $Y_{t+h}$ on the instrument $Z_{t}$, which can be thought of as the ``reduced-form.'' The denominator is the impulse response function of the assignment $W_{k,t}$ on the instrument $Z_{t}$, which can be thought of as the ``first-stage.''  
Our next result analyzes the causal interpretation of the Wald estimand in the instrumented potential outcome system. 
    
\begin{theorem}\label{thm:wald}
Assume an instrumented potential outcome system, fix $z, z^\prime \in \cW_Z$ and that   
        \begin{enumerate}
            \item[i.] \underline{Differentiability}: $Y_{t+h}(w_k)$ is continuously differentiable in the closed interval $w_k\in \mathcal{W}_k:= [\underline{w}_k, \overline{w}_k] \subset \mathbb{R}.$ 
            \item[ii.] \underline{Independence}: The instrument satisfies $Z_t \indep \{W_{k,t}(z) \colon z \in \mathcal{W}_Z\}$ and $Z_t \indep \{Y_{t+h}(w_k) \colon w_k \in \mathcal{W}_k\}$.
            \item[iii.] \underline{Relevance}: $\int_{\cW} \mathbb{E}[ 1\{W_{k,t}(z^\prime) \leq w_k \leq W_{k,t}(z)\}] dw_k>0.$
            \item[iv.] \underline{Monotonicity}: $W_{k,t}(z^\prime) \leq W_{k,t}(z)$ with probability one. 
        \end{enumerate}
    Then, Wald estimand equals, so long as it exists, 
    \[
        \frac{\int_{\cW}   \mathbb{E}[Y'_{t+h}(w_{k})|H_t(w_k)=1]   \mathbb{E}[H_t(w_k)] dw_k}
        {\int_{\cW} \mathbb{E}[H_t(w_k)] dw_k}
        ,
        \]
    where $H_t(w_k) = 1\{W_{k,t}(z^\prime) \leq w_k \leq W_{k,t}(z)\}$.
\end{theorem}

\noindent Provided the instrument is randomly assigned, relevant, and satisfies a monotonicity condition, then the Wald estimand equals a weighted average of the marginal causal effects for ``compliers'' (i.e., realizations of the potential assignment function for which moving the instrument from $z^\prime$ to $z$ changes the assignment). The marginal causal effect is the derivative of the $h$-step ahead potential outcome process with respect to the $k$-th assignment, holding all else constant. The weights are proportional to the probability of the potential assignment function being a ``complier,'' so are non-negative and sum to one.   

Since $Y_{t+h}(w_k) := Y_{t+h}(W_{1:t-1}, W_{1:k-1,t},w_{k}, W_{k+1:d_W,t}, W_{t+1:t+h})$, Assumption (ii) implicitly restricts the relationship between the instrument $Z_t$ and (1) other assignments $W_{1:k-1, 1:t+h},W_{k+1:d_W, 1:t+h}$; (2) future and past potential assignments $\{W_{k,1:t-1}(z_{1:t-1}),W_{k,t+1:t+h}(z_{t+1:t+h}) \colon z_{1:t-1}\in \mathcal{Z}^{t}, z_{t+1:t+h} \in \mathcal{Z}^{h}\}$; (3) future and past instruments $Z_{1:t-1}$ and $Z_{t+1:t+h}$; and (4) the potential outcome process $\{Y_{j,t+h}(w_{1:t+h}) \colon w_{1:t+h} \in \mathcal{W}^{t+h}\}$.
We could extend Theorem \ref{theorem: shocks, IRFs} to the instrumented potential outcome system, and show that Assumption (ii) is implied by restricting the instrument $Z_{t}$ to be independent of each of these quantities.

\begin{remark}[Binary Assignment, Binary Instrument Case]\label{remark: binary treatment, binary instrument}
Consider the simplest case with $W_{k,t} \in \{0, 1\}$, $Z_{t} \in \{0, 1\}$ and $z=1, z'=0$. 
Although the argument must be modified due to the discreteness of the assignment and instrument, we analogously show that the Wald estimand in this case equals
\[
\mathbb{E}[\{ Y_{t+h}(1)- Y_{t+h}(0) \}\mid W_{k,t}(1) - W_{k,t}(0) = 1 ].
\]
This is the time-series generalization of the binary assignment, binary instrument local average treatment effect originally derived in \cite{ImbensAngrist(94)}.
\end{remark}

\subsection{IV Estimand}
Rather than directly estimating the Wald estimand, it is natural to estimate a two-stage least squares regression of the outcome $Y_{t+h}$ on the assignment $W_{k,t}$ using the instrument $Z_{t}$. The associated IV estimand is
$$
IV_{k,t,h} := \frac{Cov(Y_{t+h},Z_t)}{Cov(W_t,Z_t)}.
$$
This has a causal interpretation by applying Theorem  \ref{theorem: local projection estimand, IRF} for the local projection estimand on both the numerator for the local projection of $Y_{t+h}$ on $Z_t$ and the denominator for the local projection of $W_{t}$ on $Z_t$. 
The statements use the notation $Y_{t+h}(z_t) := Y_{t+h}(W_{1:t-1},W_{t,1:k-1},W_k(z_t),W_{t,k+1:d_W},W_{t+1:t+h}),$ and $Y_{t+h}'(z_t) := \partial Y_{t+h}(z_t)/\partial z_t.$

\begin{theorem}\label{thm:iv}
Assume an instrumented potential outcome system. Further assume that 
\begin{enumerate}
    \item[i.] \underline{Differentiability}: $Y_{t+h}(z)$ and that $W_{t}(z)$ are continuously differentiable in the closed interval $z \in \mathcal{W}_Z = [\underline{z}, \overline{z}] \subset \mathbb{R}$.
            
    \item[ii.] \underline{Independence}: $Z_t \indep \{W_{t}(z) \colon z \in \mathcal{W}_Z\},\quad Z_t \indep \{Y_{t+h}(z) \colon z \in \mathcal{W}_Z\},$

    \item[iii.] \underline{Relevance}: $\int_{\mathcal{W}_Z} \mathbb{E}[W_{t}'(z_t)] \mathbb{E}[G_t(z_t)] d z_t \ne 0$.
\end{enumerate}

Then, it follows, if it exists, that 
$$
IV_{k,t,h} = \frac{\int_{\mathcal{W}_Z} \mathbb{E}[Y_{t+h}'(z_t)] \mathbb{E}[G_t(z_t)] d z_t}{\int_{\mathcal{W}_Z} \mathbb{E}[W_{t}'(z_t)] \mathbb{E}[G_t(z_t)] d z_t}
$$
where $G_t(z_t) = 1\{z_t \leq Z_t\}(Z_t-\mathbb{E}[Z_t])$, noting $\mathbb{E}[G_t(z_t)] \ge 0$.
\end{theorem}

\subsection{Generalized Wald Estimand}

The generalized Wald estimand is a ratio of a reduced-form generalized impulse response function to a first-stage generalized impulse response function. It is given by, for fixed $z, z^\prime \in \cW_Z$, 
\begin{equation}
    \frac{ \mathbb{E}[Y_{t+h} \mid Z_{t} = z, \mathcal{F}_{t-1}] - \mathbb{E}[ Y_{t+h} \mid Z_{t} = z^\prime, \mathcal{F}_{t-1} ] }{ \mathbb{E}[W_{k,t} \mid Z_{t} = z, \mathcal{F}_{t-1}] - \mathbb{E}[ W_{k, t} \mid Z_{t} = z^\prime, \mathcal{F}_{t-1}] }.
\end{equation}

\begin{theorem}\label{theorem: instrument, observable treatments}
    Assume an instrumented potential outcome system, fix $z, z^\prime \in \cW_Z$ and that    
        \begin{enumerate}
            \item[i.] \underline{Differentiability}: $Y_{t+h}(w_k)$ is continuously differentiable in the closed interval $w_k\in \mathcal{W}_k:= [\underline{w}_k, \overline{w}_k] \subset \mathbb{R}.$ 
        
            \item[ii.] \underline{Independence}: The instrument satisfies $[Z_t \indep \{W_{k,t}(z) \colon z \in \mathcal{W}_Z\}] \mid \mathcal{F}_{t-1}$ and $[Z_t \indep \{Y_{t+h}(w_k) \colon w_k \in \mathcal{W}_k\}] \mid \mathcal{F}_{t-1}$.
            
            \item[iii.] \underline{Relevance}: $\int_{\cW} \mathbb{E}[ 1\{W_{k,t}(z^\prime) \leq w_k \leq W_{k,t}(z)\} \mid \mathcal{F}_{t-1} ] dw_k>0.$ 
        
            \item[iv.] \underline{Monotonicity}: $W_{k,t}(z^\prime) \leq W_{k,t}(z)$ with probability one. 
        \end{enumerate}
    Then, the generalized Wald estimand equals, so long as it exists, 
    \[
        \frac{\int_{\cW}   \mathbb{E}[Y'_{t+h}(w_{k}) \mid H_t(w_k)=1,\mathcal{F}_{t-1}]   \mathbb{E}[H_t(w_k)\mid \mathcal{F}_{t-1}] dw_k}
        {\int_{\cW} \mathbb{E}[H_t(w_k) \mid \mathcal{F}_{t-1}] dw_k}
        ,
        \]
        where, again, $H_t(w_k) = 1\{W_{k,t}(z^\prime) \leq w_k \leq W_{k,t}(z)\}$.
\end{theorem}

The generalized Wald estimand analogously equals a weighted average of the marginal filtered causal effects for ``compliers,'' where the weights are proportional to the probability of the potential assignment function being a ``complier'' conditional on the filtration.

We next provide a sufficient condition for the instrument to be randomly assigned in terms of conditional independence restrictions on these underlying processes.
\begin{theorem}\label{theorem: instrument random assignment suff cond}
    Assume that the instrument satisfies $Z_t \indep (Z_{t+1:t+h}, W_{1:k-1, t:t+h}, \{W_{k,t+1:t+h}(z_{t+1:t+h}) \colon z_{t+1:t+h} \in \mathcal{Z}^{h}\}, W_{k+1:d_W, t:t+h}, \{Y_{t+h}(w_{1:t+h}) \colon w_{1:t+h} \in \mathcal{W}^{t+h}\} ) \mid \mathcal{F}_{t-1}$.
    Then, Assumption (ii) in Theorem \ref{theorem: instrument, observable treatments} is satisfied.
\end{theorem}

\subsection{Generalized IV and Filtered IV Estimands}

Estimating the generalized Wald estimand is not easy, particularly if $Z_t$ is not discrete. Here we derive a causal interpretation for \textit{generalized IV estimand} 
$$
\frac{Cov(Y_{t+h},Z_t \mid \mathcal{F}_{t-1}) }{ Cov(W_{k,t},Z_t\mid \mathcal{F}_{t-1}) } = \frac{\mathbb{E}[(Y_{t+h} - \hat{Y}_{t+h \mid t - 1})(Z_{t} - \hat{Z}_{t \mid t - 1})\mid \mathcal{F}_{t-1}]}{\mathbb{E}[(W_{k,t} - \hat{W}_{k,t \mid t - 1})(Z_{t} - \hat{Z}_{t\mid t-1})\mid \mathcal{F}_{t-1}]}.
$$
where $\hat{Y}_{t+h \mid t - 1} = \mathbb{E}[Y_{t+h}\mid \mathcal{F}_{t-1}]$, $\hat{W}_{k,t\mid t - 1} = \mathbb{E}[W_{k,t}\mid \mathcal{F}_{t-1}]$ and $\hat{Z}_{t \mid t - 1} = \mathbb{E}[Z_{t}\mid \mathcal{F}_{t-1}]$. 
No new technical issues arise in dealing with this setup, but Assumption (ii) in Theorem \ref{thm:iv} now becomes 
\begin{equation}\label{eqn:condind}
[Z_t \indep \{Y_{t+h}(z) \colon z \in \mathcal{W}_Z\}] \mid \mathcal{F}_{t-1},\quad 
[Z_t \indep \{W_{t}(z) \colon z \in \mathcal{W}_Z\}]\mid \mathcal{F}_{t-1}.
\end{equation}
Then, the generalized IV estimand equals 
$$\frac{\int_{\mathcal{W}_Z} \mathbb{E}[Y_{t+h}'(z_t)\mid \mathcal{F}_{t-1}] \mathbb{E}[G_{t|t-1}(z_t)\mid \mathcal{F}_{t-1}] d z_t}{\int_{\mathcal{W}_Z} \mathbb{E}[W_{t}'(z_t)\mid \mathcal{F}_{t-1}] \mathbb{E}[G_{t|t-1}(z_t)\mid \mathcal{F}_{t-1}] d z_t}
$$
where $G_{t|t-1}(z_t) = 1\{z_t \le Z_t\}(Z_t-\mathbb{E}[Z_t\mid \mathcal{F}_{t-1}])$, noting $\mathbb{E}[G_{t|t-1}(z_t)\mid \mathcal{F}_{t-1}] \ge 0$.

Of more practical importance is the \textit{filtered IV estimand}
$$
\frac{\mathbb{E}[(Y_{t+h} - \hat{Y}_{t+h \mid t-1})(Z_{t} - \hat{Z}_{t\mid t-1})]}{\mathbb{E}[(W_{k,t} - \hat{W}_{k,t \mid t-1})(Z_{t} - \hat{Z}_{t \mid t-1})],
},
$$
which can be estimated by instrumental variables applied to $Y_{t+h} - \hat{Y}_{t+h \mid t-1}$ on $W_{k,t} - \hat{W}_{k,t \mid t-1}$ with instruments $Z_t - \hat{Z}_{t\mid t - 1}$. Under the conditions of Theorem \ref{thm:iv} but using Equation (\ref{eqn:condind}) instead of Assumption (ii), then the filtered IV estimand becomes 
$$\frac{\int_{\mathcal{W}_Z} \mathbb{E}[\mathbb{E}[Y_{t+h}'(z_t)\mid \mathcal{F}_{t-1}] \mathbb{E}[G_{t|t-1}(z_t)\mid \mathcal{F}_{t-1}]] d z_t}{\int_{\mathcal{W}_Z} \mathbb{E}[\mathbb{E}[W_{t}'(z_t)\mid \mathcal{F}_{t-1}] \mathbb{E}[G_{t|t-1}(z_t)\mid \mathcal{F}_{t-1}]] d z_t}.
$$

\section{Estimands Based on Instruments and Outcomes}\label{sect:instrOutcome}

In this section, we study conditions under which common statistical estimands based on only instruments and outcomes have causal meaning. 
We focus on an instrumented potential outcome system $$\{Z_t,\{W_t(z_t),z_t\in \mathcal{W}_Z\},\{Y_t(w_{1:t}),w_{1:t}\in \mathcal{W}_W^t\}\}_{t \ge 1},$$ in which the researcher only observes the instruments and the outcomes $\{z_t^{obs},y_t^{obs}\}_{t\ge 1}$. 
We will sometimes refer to $\{\mathcal{F}_{t}^{Z,Y}\}_{t \ge 1}$ as the natural filtration generated by the realized $\{z_t^{obs},y_t^{obs}\}_{t \ge 1}$. 

In this context, it is common for empirical researchers to analyze estimands involving two elements of the outcome vector $Y_{j,t+h}$, $Y_{k,t}$ and the instrument $Z_{t}$ (therefore, we return to using a explicit subscript on the outcome variable). 
Consider, for example, an empirical researcher that constructs an instrument $Z_{t}$ for the monetary policy shock (e.g., an instrument of the form used in \cite{Kuttner(01), CochranePiazessi(02), GertlerKaradi(15)} or \cite{RomerRomer(04)}). 
In this case, the empirical researcher may measure the dynamic causal effect of the monetary policy shock $W_{k,t}$ on unemployment $Y_{j,t+h}$ by estimating the first-stage impulse response function of the federal funds rate $Y_{k,t}$ on the instrument $Z_{t}$. See, for example, \cite{JordaSchularickTaylor(15), RameyZubairy(18), JordaSchularickTaylor(20)} for recent applications of this empirical strategy. 

In particular, we study the following estimands: Ratio Wald, Local Projection IV, generalized Ratio Wald, and the local filtered projection IV.
We show that such dynamic IV estimands identify ``relative'' local average impulse causal effect, which is a generalization of the interpretation of such a dynamic IV estimand in existing literature on external instruments \citep{StockWatson(18), PlagborgMollerWolf(20), JordaSchularickTaylor(20)}.
Table \ref{tab:instrumentsoutcome} defines these estimands and summarizes our main results on their causal interpretation under important restrictions on the assignment process and other technical conditions. 

\begin{table}[htbp!]
{\footnotesize
   \centering
    \begin{tabular}{l || l|l}
    \textbf{Name}  & \textbf{Estimand} & \textbf{Causal Interpretation} \\ \hline \hline
        Ratio Wald &$\frac{ \mathbb{E}[Y_{j,t+h} \mid Z_{t} = z] - \mathbb{E}[ Y_{j,t+h} \mid Z_{t} = z^\prime] }{ \mathbb{E}[Y_{k,t} \mid Z_{t} = z] - \mathbb{E}[ Y_{k, t} \mid Z_{t} = z^\prime] }$& $\frac{\int_{\cW}  \mathbb{E}[Y'_{j,t+h}(w_k) \mid H_t(w_k)=1]   \mathbb{E}[H_t(w_k)] dw_k}
        {\int_{\cW}   \mathbb{E}[Y'_{k,t}(w_k) \mid H_t(w_k)=1]   \mathbb{E}[H_t(w_k)] dw_k}$ \\
        && \\
        && \\
        Local Projection& $\frac{ Cov(Y_{j,t+h}, Z_{t}) }{ Cov(Y_{k,t}, Z_{t})}$ & $\frac{\int_{\mathcal{W}_Z} \mathbb{E}[Y'_{j,t+h}(z_{k})] \mathbb{E}[G_t(z_k)] dz_k}{\int_{\mathcal{W}_Z} \mathbb{E}[Y'_{k,t}(z_{k})] \mathbb{E}[G_t(z_k)] dz_k}$  \\
        IV&& \\
        && \\
        Generalized Ratio & $\frac{ \mathbb{E}[Y_{j,t+h} \mid Z_{t} = z, \mathcal{F}_{t-1}^{Z,Y}] - \mathbb{E}[ Y_{j,t+h} \mid Z_{t} = z^\prime, \mathcal{F}_{t-1}^{Z,Y} ] }{ \mathbb{E}[Y_{k,t} \mid Z_{t} = z, \mathcal{F}_{t-1}^{Z,Y} ] - \mathbb{E}[ Y_{k, t} \mid Z_{t} = z^\prime, \mathcal{F}_{t-1}^{Z,Y} ] }$       
        & $\frac{\int_{\cW}   \mathbb{E}[Y'_{j,t+h}(w_k) \mid H_t(w_k)=1,\mathcal{F}_{t-1}^{Z,Y}]   \mathbb{E}[H_t(w_k) \mid \mathcal{F}_{t-1}^{Z,Y}] dw_k}
        {\int_{\cW}   \mathbb{E}[Y'_{k,t}(w_k) \mid H_t(w_k)=1,\mathcal{F}_{t-1}^{Z,Y}]   \mathbb{E}[H_t(w_k)\mid \mathcal{F}_{t-1}^{Z,Y}] dw_k}$ \\
        Wald && \\
        && \\
%        Generalized Local  &$\frac{ Cov(Y_{j,t+h}, Z_{t} \mid \mathcal{F}_{t-1}^{Z,Y}) }{ Cov(Y_{k,t}, Z_{t}\mid \mathcal{F}_{t-1}^{Z,Y})  }$& $\frac{\int_{\mathcal{W}_Z} \mathbb{E}[Y'_{j,t+h}(z_{k})\mid \mathcal{F}_{t-1}^{Z,Y}] \mathbb{E}[G_t(z_k)\mid \mathcal{F}_{t-1}^{Z,Y}] dz_k}{\int_{\mathcal{W}_Z} \mathbb{E}[Y'_{k,t}(z_{k})\mid \mathcal{F}_{t-1}^{Z,Y}] \mathbb{E}[G_t(z_k)\mid \mathcal{F}_{t-1}^{Z,Y}] dz_k}$ \\
%        Projection IV && \\
%        && \\
        Local Filtered  &$\frac{ Cov(Y_{j,t+h}-\hat{Y}_{j,t+h\mid t - 1},Z_t- \hat{Z}_{t\mid t - 1}) }{ Cov(Y_{k,t}-\hat{Y}_{k,t\mid t - 1}, Z_t - \hat{Z}_{t\mid t - 1})  }$& $\frac{\int_{\mathcal{W}_Z} \mathbb{E}[\mathbb{E}[Y'_{j,t+h}(z_{k})\mid \mathcal{F}_{t-1}^{Z,Y}] \mathbb{E}[G_t(z_k)\mid \mathcal{F}_{t-1}^{Z,Y}]] dz_k}{\int_{\mathcal{W}_Z} \mathbb{E}[\mathbb{E}[Y'_{k,t}(z_{k})\mid \mathcal{F}_{t-1}^{Z,Y}] \mathbb{E}[G_t(z_k)\mid \mathcal{F}_{t-1}^{Z,Y}]] dz_k}$ \\
        Projection IV&& 
    \end{tabular}
    \caption{Top line results for the causal interpretation of common estimands based on instruments and outcomes.  Here $H_t(w_k) = 1\{W_{k,t}(z^\prime) \leq w_k \leq W_{k,t}(z)\}$, $G_t(z_t) = 1\{z_t \le Z_t\}(Z_t-\mathbb{E}[Z_t])$ and $G_{t|t-1}(z_t) = 1\{z_t \le Z_t\}(Z_t-\mathbb{E}[Z_t \mid \mathcal{F}_{t-1}^{Z,Y})$, while $\hat{Y}_{k,t+h|t-1} = \mathbb{E}[Y_{k,t+h}\mid \mathcal{F}_{t-1}^{Z,Y}]$ and $\hat{Z}_{t\mid t - 1} = \mathbb{E}[Z_t\mid \mathcal{F}_{t-1}^{Z,Y}]$.  Note that $\mathbb{E}[G_t(z_t)] \ge 0$ and $\mathbb{E}[G_{t|t-1}(z_t)\mid \mathcal{F}_{t-1}^{Z,Y}] \ge 0$.}
   \label{tab:instrumentsoutcome}
    }
\end{table}

\subsection{Ratio Wald Estimand}

The \textit{Ratio Wald Estimand}
    $$
    \frac{ \mathbb{E}[Y_{j,t+h} \mid Z_{t} = z] - \mathbb{E}[ Y_{j,t+h} \mid Z_{t} = z^\prime] }{ \mathbb{E}[Y_{k,t} \mid Z_{t} = z] - \mathbb{E}[ Y_{k, t} \mid Z_{t} = z^\prime] },
    $$
which is the ratio of the Wald estimands:
\[
    \frac{ \mathbb{E}[Y_{j,t+h} \mid Z_{t} = z] - \mathbb{E}[ Y_{j,t+h} \mid Z_{t} = z^\prime ] }{ \mathbb{E}[W_{k,t} \mid Z_{t} = z] - \mathbb{E}[ W_{k, t} \mid Z_{t} = z^\prime] },
    \quad \text{to} \quad 
    \frac{ \mathbb{E}[Y_{k,t} \mid Z_{t} = z] - \mathbb{E}[ Y_{k,t} \mid Z_{t} = z^\prime ] }{ \mathbb{E}[W_{k,t} \mid Z_{t} = z] - \mathbb{E}[ W_{k, t} \mid Z_{t} = z^\prime] }.
    \]
Hence we just need to collect the conditions for the validity of their causal representations, and then apply Theorem \ref{thm:wald} twice.

\begin{corollary}\label{corollary: ratio wald}
Assume an instrumented potential outcome system, $z, z^\prime \in \mathcal{W}_Z$ and that   
    \begin{enumerate}
        \item[i.] \underline{Differentiability}: $Y_{k,t}(w_k),Y_{j,t+h}(w_k)$ are continuously differentiable in closed interval $\mathcal{W}_k:= [\underline{w}_k, \overline{w}_k] \subset \mathbb{R}.$ 
        
        \item[ii.] \underline{Independence}: $Z_t \indep \{W_{k,t}(z) \colon z \in \mathcal{W}_Z\}$ and $Z_t \indep \{Y_{k,t}(w_k),Y_{j,t+h}(w_k) \colon w_k \in \mathcal{W}_k\}$.
        
        \item[iii.] \underline{Relevance}: $\int_{\cW} \mathbb{E}[Y'_{k,t}(w_k) \mid H_t(w_k)=1]   \mathbb{E}[H_t(w_k)] dw_k \ne 0.$    
            
        \item[iv.] \underline{Monotonicity}: $W_{k,t}(z^\prime) \leq W_{k,t}(z)$ with probability one. 
        \end{enumerate}
Then, the Ratio Wald Estimand equals, if it exists,
 \[
    \frac{\int_{\cW}  \mathbb{E}[Y'_{j,t+h}(w_k) \mid H_t(w_k)=1]   \mathbb{E}[H_t(w_k)] dw_k}
    {\int_{\cW}   \mathbb{E}[Y'_{k,t}(w_k) \mid H_t(w_k)=1]   \mathbb{E}[H_t(w_k)] dw_k},
\]
where $H_t(w_k) = 1\{W_{k,t}(z^\prime) \leq w_k \leq W_{k,t}(z)\}$.
\end{corollary}

In words, the ratio Wald estimand above identifies a \textit{relative} local average impulse causal effect under the instrumented potential outcome system. 
The numerator is a weighted average of the marginal causal effects of $W_{k,t}$ on the $h$-step ahead outcome $Y_{j,t+h}$, where the weights are proportional to the probability of compliance. Similarly, the denominator is a weighted average of the marginal causal effects of $W_{k,t}$ on the contemporaneous outcome $Y_{k,t}$. 
Therefore, the ratio in Corollary \ref{corollary: ratio wald} measures the causal response of the $h$-step ahead outcome $Y_{j,t+h}$ to a change in the treatment $W_{k,t}$ that increases the contemporaneous outcome $Y_{k,t}$ by one unit on impact (among compliers). 

This is a generalization of the well-known result that in linear SVMA models (without invertibility) the IV based estimands identify relative impulse response functions \citep{StockWatson(18),PlagborgMollerWolf(20)}. Corollary \ref{corollary: ratio wald} makes no functional form assumptions nor standard time series assumptions such as invertibility or recoverability. 
In this sense, Corollary \ref{corollary: ratio wald} highlights the attractiveness of using external instruments to measure dynamic causal effects in observational time series data. 
Provided there exists an external instrument for the treatment $W_{k,t}$ that is randomly assigned, relevant and satisfies a monotonicity condition, then the researcher can identify causally interpretable estimands without further assumptions and without even directly observing the treatment itself. 

\subsection{Local Projection IV Estimand}

The \textit{local projection IV estimand}
    $$
    \frac{ Cov(Y_{j,t+h}, Z_{t}) }{ Cov(Y_{k,t}, Z_{t})},
    $$
is the ratio of the IV estimands $\frac{ Cov(Y_{j,t+h}, Z_{t}) }{ Cov(W_{k,t}, Z_{t})}$ to $\frac{ Cov(Y_{k,t}, Z_{t}) }{ Cov(W_{k,t}, Z_{t})}$.
Therefore, we once again just need to collect the conditions for the validity of their causal representations, and apply Theorem \ref{thm:iv} twice.

\begin{corollary}\label{cor:LPIV} Consider an instrumented potential outcome system.  Further assume that 
\begin{enumerate}
    \item[i.] \underline{Differentiability}: $Y_{k,t}(z),Y_{j,t+h}(z),W_{t}(z)$ are continuously differentiable in the closed interval $z \in \mathcal{W}_Z = [\underline{z}, \overline{z}] \subset \mathbb{R}$.
            
    \item[ii.] \underline{Independence}: $Z_t \indep \{Y_{k,t}(z),Y_{j,t+h}(z) \colon z \in \mathcal{W}_Z\}$ and $Z_t \indep \{W_{t}(z) \colon z \in \mathcal{W}_Z\}$.

    \item[iii.] \underline{Relevance}: $\int_{\mathcal{W}_Z} \mathbb{E}[Y_{k,t}'(z_t)] \mathbb{E}[G_t(z_t)] d z_t \ne0$.
\end{enumerate}

Then, the local projection IV estimand equals  
$$ \frac{\int_{\mathcal{W}_Z} \mathbb{E}[Y'_{j,t+h}(z_{k})] \mathbb{E}[G_t(z_k)] dz_k}{\int_{\mathcal{W}_Z} \mathbb{E}[Y'_{k,t}(z_{k})] \mathbb{E}[G_t(z_k)] dz_k},
$$
where $G_t(z_k)= 1\{z_k \leq Z_{t}\} (Z_{t} - \mathbb{E}[Z_{t}])$, noting $\mathbb{E}[G_t(z_k)] \ge 0$.
\end{corollary}

\subsection{Generalized Ratio Wald Estimand}

Researchers may also be interested in analyzing the \textit{generalized ratio Wald estimand}:
    \[
    \frac{ \mathbb{E}[Y_{j,t+h} \mid Z_{t} = z, \mathcal{F}_{t-1}^{Z,Y}] - \mathbb{E}[ Y_{j,t+h} \mid Z_{t} = z^\prime, \mathcal{F}_{t-1}^{Z,Y} ] }{ \mathbb{E}[Y_{k,t} \mid Z_{t} = z, \mathcal{F}_{t-1}^{Z,Y} ] - \mathbb{E}[ Y_{k, t} \mid Z_{t} = z^\prime, \mathcal{F}_{t-1}^{Z,Y} ] },
    \]
which is the ratio of generalized impulse response functions at different lags and for different outcome variables. 
Since this is the ratio of two generalized Wald estimands, we immediately arrive at the following corollary by applying Theorem \ref{theorem: instrument, observable treatments} twice.

\begin{corollary}\label{corr: IV with unobservable treatments}
   Assume an instrumented potential outcome system, $z, z^\prime \in \mathcal{W}_Z$ and that 
        \begin{enumerate}
            \item[i.] \underline{Differentiability}: $Y_{k,t}(w_k),Y_{j,t+h}(w_k)$ are continuously differentiable in closed interval $\mathcal{W}_k:= [\underline{w}_k, \overline{w}_k] \subset \mathbb{R}.$ 
            
            \item[ii.] \underline{Independence}: $[Z_t \indep \{W_{k,t}(z) \colon z \in \mathcal{W}_Z\}]\mid \mathcal{F}_{t-1}^{Z,Y}$ and $[Z_t \indep \{Y_{k,t}(w_k),Y_{j,t+h}(w_k) \colon w_k \in \mathcal{W}_k\}]\mid \mathcal{F}_{t-1}^{Z,Y}$.
            
            \item[iii.] \underline{Relevance}: $\int_{\cW} \mathbb{E}[Y'_{k,t}(w_k) \mid H_t(w_k)=1, \mathcal{F}_{t-1}^{Z,Y}]   \mathbb{E}[H_t(w_k)\mid \mathcal{F}_{t-1}^{Z,Y}] dw_k \ne 0.$ 
            
            \item[iv.] \underline{Monotonicity}: $W_{k,t}(z^\prime) \leq W_{k,t}(z) \mid \mathcal{F}_{t-1}^{Z,Y}$ with probability one. 
        \end{enumerate}
    Then, the generalized ratio Wald estimand equals
         \[
        \frac{\int_{\cW}   \mathbb{E}[Y'_{k,t+h}(w_k) \mid H_t(w_k)=1,\mathcal{F}_{t-1}^{Z,Y}]   \mathbb{E}[H_t(w_k)\mid \mathcal{F}_{t-1}^{Z,Y}] dw_k}
        {\int_{\cW}   \mathbb{E}[Y'_{k,t}(w_k) \mid H_t(w_k)=1,\mathcal{F}_{t-1}^{Z,Y}]   \mathbb{E}[H_t(w_k)\mid \mathcal{F}_{t-1}^{Z,Y}] dw_k},
        \]
        where $H_t(w_k) = 1\{W_{k,t}(z^\prime) \leq w_k \leq W_{k,t}(z)\}$.
\end{corollary}

\noindent The interpretation of Corollary \ref{corr: IV with unobservable treatments} is analogous to the interpretation of the ratio Wald estimand in Corollary \ref{corollary: ratio wald}, except now everything is conditional on the natural filtration.

\subsection{Generalized Local Projection IV and Local Filtered Projection IV Estimands}
In practice, researchers typically estimate generalized impulse response functions using a two-stage least-squares type estimator. This is also sometimes called ``local projections with an external instrument'' \citep{JordaSchularickTaylor(15)}. We first analyze this \textit{generalized local projection IV}
\begin{equation}
    \frac{ Cov(Y_{j,t+h}, Z_{t} \mid \mathcal{F}_{t-1}^{Z,Y}) }{ Cov(Y_{k,t}, Z_{t}  \mid \mathcal{F}_{t-1}^{Z,Y})},
\end{equation}
which again is a ratio, this time of the Generalized IV estimands at different lag lengths.  Using the same arguments as Corollary \ref{cor:LPIV}, it has the causal interpretation 
$$ \frac{\int_{\mathcal{W}_Z} \mathbb{E}[Y'_{j,t+h}(z_{k})\mid \mathcal{F}_{t-1}^{Z,Y}] \mathbb{E}[G_t(z_k)\mid\mathcal{F}_{t-1}^{Z,Y}] dz_k}{\int_{\mathcal{W}_Z} \mathbb{E}[Y'_{k,t}(z_{k})\mid \mathcal{F}_{t-1}^{Z,Y}] \mathbb{E}[G_t(z_k)\mid \mathcal{F}_{t-1}^{Z,Y}] dz_k},
$$
where $G_{t|t-1}(z_k)= 1\{z_k \leq Z_{t}\} (Z_{t} - \mathbb{E}[Z_{t}\mid \mathcal{F}_{t-1}^{Z,Y}])$.

Of more practical relevance, is the \textit{local filtered projection IV estimand} is 
$$
    \frac{ Cov(Y_{j,t+h} - \hat{Y}_{j,t+h \mid t - 1},Z_t - \hat{Z}_{t \mid t - 1}) }{ Cov(Y_{k,t}-\hat{Y}_{k,t \mid t - 1}, Z_t-\hat{Z}_{t \mid t - 1})  },
    $$
where recall that, for example, $\hat{Y}_{k,t+h} = \mathbb{E}[Y_{k, t+h} \mid \mathcal{F}_{t-1}^{Z,Y}]$, and $\hat{Z}_{t \mid t-1 } = \mathbb{E}[Z_{t}\mid \mathcal{F}_{t-1}^{Z,Y}]$.
The properties of this are inherited from those of the generalized local projection IV.  
In particular, it equals 
$$ \frac{\int_{\mathcal{W}_Z} \mathbb{E}[\mathbb{E}[Y'_{j,t+h}(z_{k})\mid \mathcal{F}_{t-1}^{Z,Y}] \mathbb{E}[G_t(z_k)\mid \mathcal{F}_{t-1}^{Z,Y}]] dz_k}{\int_{\mathcal{W}_Z} \mathbb{E}[\mathbb{E}[Y'_{k,t}(z_{k})\mid \mathcal{F}_{t-1}^{Z,Y}] \mathbb{E}[G_t(z_k)\mid \mathcal{F}_{t-1}^{Z,Y}]] dz_k}.
$$
%where, again, $G_{t|t-1}(z_k)= 1\{z_k \leq Z_{t}\} (Z_{t} - \mathbb{E}[Z_{t}\mid \mathcal{F}_{t-1}^{Z,Y}])$.

\section{Estimands Based Only on Outcomes}\label{ref:onlyoutcomes}

The dominant approach to causal inference in macroeconometrics is a model-based approach in the tradition of \cite{Sims(80)}. 
See, for example, \cite{Ramey(16)} and \cite{KilianLutkepohl(17)} for recent reviews.
In that literature, researchers introduce parametric models to study the dynamic causal effects of unobservable ``structural shocks,'' which themselves must be inferred from the outcomes. 
Here we link this to our setup, mostly to place our work in context and illustrate that simultaneous equation modeling can be nested in the direct potential outcome system framework.
Assume there is a direct potential outcome system $\{W_t,\{Y_t(w_{1:t}):w_{1:t} \in \mathcal{W}_w^t\}_{t\ge 1}$, where researchers only see the outcomes $\{y_t^{obs}\}_{t \ge 1}$.

\subsection{Linear Simultaneous Equation Approach}

The causal inference approach of using only time series data on outcomes is in the storied tradition of linear simultaneous equations models developed at the Cowles Foundation \citep[e.g.,][]{Christ(94), Hausman(83)}. 
The most essential causal challenges arise without any dynamic causal effects, so we start with a static example as an illustration. Suppose that
\[
A_{0}Y_{t}(w_{1:t})=\alpha + w_{t},\quad w_{1:t} \in \mathcal{W}^{t},\quad
t=1,2,...,
\]%
where $A_{0}$ is a non-stochastic, square matrix. Notice that in this model the potential outcome process is deterministic and linear combinations of the potential outcomes equal the possible assignments for every $t$. If $A_{0}$ is additionally invertible, then $Y_{t}(w_{1:t})=A_{0}^{-1}\left( \alpha +w_{t}\right),$ which implies that the contemporaneous average treatment effect is $\mathbb{E}[Y_{t}(W_{1:t-1},w)-Y_{t}(W_{1:t-1},w^{\prime })] = A_{0}^{-1}\left( w-w^{\prime }\right)$, and the marginal average treatment effect is $\mathbb{E}[\frac{\partial
Y_{t}(w_{1:t})}{\partial w_{t}^{\mathrm{T}}}]=A_{0}^{-1}$ whatever probabilistic assumption is made about $W_{1:t-1}$

Furthermore, under this model, if the second moments of the observables exist and $\mathrm{Var}(W_{t})$ is non-singular, then $\mathrm{Cov}(Y_{t},W_{t})\mathrm{Var}(W_{t})^{-1}=A_{0}^{-1}$ for every $t$, which would make statistical inference rather straightforward. 
But the point of this simultaneous equations literature is to carry out inference without directly observing the assignments --- which is a much harder task. 

Notice further that under the model $\mathrm{Var}(Y_{t})=A_{0}^{-1}\mathrm{Var}(W_{t})\left( A_{0}^{-1}\right)^{\mathrm{T}}$.
Identifying $\mathrm{Var}(Y_{t})$ is not enough to untangle $A_{0}$ and $\mathrm{Var}(W_{t})$, and so knowledge of the second moments of the observables is not enough alone to learn the contemporaneous average treatment effect. 
In the linear simultaneous equations literature, this is resolved by imposing more economic structure on the potential outcome process, such as placing more structure on the matrix $A_0$.

A central a priori constraint is the one highlighted by \cite{Sims(80)}. He imposed that (a) $A_{0}$ is triangular, (b) $\mathrm{Var}(W_{t})$ is diagonal. For simplicity of exposition, look at the two dimensional case and write $A_{0}=\left( 
\begin{array}{cc}
1 & 0 \\ 
-a_{21} & 1%
\end{array}%
\right)$ and $\mathrm{Var}(W_{t})=\left( 
\begin{array}{cc}
\sigma _{11}^{2} & 0 \\ 
0 & \sigma _{22}^{2}%
\end{array}%
\right)$. Then the elements within $A_{0}$ and $\mathrm{Var}(W_{t})$ can be individually determined from $\mathrm{Var}(Y_{t})$ if $\mathrm{Var}(Y_{t})$ is of full rank. The same holds in higher dimensions. Hence, with additional restrictions on the potential outcome process, the contemporaneous causal effect can be determined from the data on the outcomes, without having observing the assignments (or without the access to instruments). There are of course many other a priori constraints that could be imposed and the above structure extends to non-linear systems of equations $g\left(Y_{t}(w_{1:t})\right)=w_{t}$.

The linear \textquotedblleft structural vector autoregression \textquotedblright\ (SVAR) version of the linear simultaneous equation has the same fundamental structure. Focusing on the one lag model with no intercept for simplicity, the SVAR approach assumes that the potential outcome process satisfies $A_{0}Y_{t}(w_{1:t})=w_{t}+A_{1}Y_{t-1}(w_{1:t-1})$.
\cite{KilianLutkepohl(17)} provide a book length review of this model structure and its various extensions and implications. 
Then $A_{0}\left( I-\Phi _{1}L\right) Y_{t}(w_{1:t})=w_{t}$, where $L$ is a lag operator and $\Phi _{1}=A_{0}^{-1}A_{1}$. So $Y_{t}(w_{1:t})=A_{0}^{-1}w_{t}+\Phi _{1}Y_{t-1}(w_{1:t-1})$, which in turn implies that the potential outcome process also has an SVMA model representation
\[
Y_{t}(w_{1:t})=A_{0}^{-1}w_{t}+\Phi _{1}A_{0}^{-1}w_{t-1}+\Phi
_{1}^{2}A_{0}^{-1}w_{t-2}+...+\Phi _{1}^{t-1}A_{0}^{-1}w_{1}+\Phi
_{1}^{t}A_{0}^{-1}Y_{0}.
\]%
In this case, the $h$-period ahead average treatment effect is $$\mathbb{E}[Y_{t+h}(W_{1:t-1},w,W_{t+1:t+h})-Y_{t+h}(W_{1:t-1},w^{\prime
},W_{t+1:t+h})]=\Phi _{1}^{h}A_{0}^{-1}\left( w-w^{\prime }\right)$$ and the $h$-period ahead marginal average treatment effect is $\mathbb{E}[ \partial Y_{t+h}(w_{1:t+h})/\partial w_{t}^{\prime } ] = \Phi _{1}^{h}A_{0}^{-1}$.
The time series parameter $\Phi_{1}$ can be determined from the dynamics of
the observable outcomes if this process is stationary. But again $A_{0}$ and $\mathrm{Var}(W_{t})$ cannot be separately identified from the observable outcomes, so further assumptions are needed. 

\subsection{Causal Interpretation of the GIRF of $Y_{k,t}$ on $Y_{j,t+h}$}

A broader analysis focuses on the $h$-step ahead generalized impulse response function of the $j$-th outcome on the $k$-th outcome without placing functional form restrictions on the potential outcome process. Here we provide a nonparametric causal meaning to it in terms of potential outcomes.  To do so, we will further assume that the potential outcome process is a deterministic function of the assignments and that the assignments are independent across time.
%We will give conditions under which the GIRF equals 
%$$
%\mathrm{E}[\psi_{j,t+h}(W_{1:t})|Y_{k,t}=y_k,\mathcal{F}^Y_{t-1}] 
%- \mathrm{E}[\psi_{j,t+h}(W_{1:t})|Y_{k,t}=y_k',\mathcal{F}^Y_{t-1}],
%$$
%which relates to the ``stochastic treatment effect'' associated with  \cite{Stock(09)}, \cite{MunozvanderLaan(12)} and \cite{Papadogeorgou(21)}.

% \begin{assumption}[Deterministic potential outcome process]\label{ass:degenerate} 
% Assume the potential outcome process $Y_{t}(w_{1:t})$ is deterministic for all $t \geq 0$, $w_{1:t} \in \mathcal{W}^t$. 
%\end{assumption}

%\begin{assumption}\label{ass:condInd} Assume that for all $t\ne s$, the $ W_t \indep W_s.$
%%(b) that $W_{k,t} \indep W_{j,t}$ for all $j\ne k$.
%\end{assumption}

\begin{theorem}\label{thm:deterPO}  Consider a direct potential outcome system, and further assume that
\begin{enumerate}
    \item[i.] the potential outcome process $Y_{t}(w_{1:t})$ is deterministic for all $t \geq 0$, $w_{1:t} \in \mathcal{W}^t$.
    \item[ii.] for all $t\ne s$, $ W_t \indep W_s$.
\end{enumerate}
Then, so long as the corresponding moments exist,
\begin{align}
&\mathrm{E}[Y_{j,t+h} \mid Y_{k,t}=y_k,\mathcal{F}^Y_{t-1}] - \mathrm{E}[Y_{j,t+h} \mid Y_{k,t}=y_k',\mathcal{F}^Y_{t-1}] \\
& = 
\mathrm{E}[\psi_{j,t+h}(W_{1:t}) \mid Y_{k,t}=y_k,\mathcal{F}^Y_{t-1}] 
- \mathrm{E}[\psi_{j,t+h}(W_{1:t})\mid Y_{k,t}=y_k',\mathcal{F}^Y_{t-1}], \label{eqn:psi1}
\end{align}
where 
$
\psi_{j,t+h}(w_{1:t}) := \mathrm{E}[Y_{j,t+h}(w_{1:t},W_{t+1:t+h})].
$
\end{theorem}
% \noindent Proof.  Given in the Appendix. 

\noindent Theorem \ref{thm:deterPO} illustrates that without functional form restrictions on the potential outcome process, the generalized impulse response function of the $j$-th outcome on the $k$-th outcome has a causal interpretation in terms of the shifting the entire conditional distribution of the treatments $W_{1:t}$.
While this is a non-standard object, it can be interpreted as the causal effect of a stochastic intervention on the assignment path $W_{1:t}$, which has been an object of recent interest in a growing cross-sectional literature on causal inference in the presence of interference/spillovers across units -- see, for example, \cite{MunozvanderLaan(12)}, \cite{PapadogeorgouMealliZigler(19)}, \cite{PapadogeorgouEtAl(21)}, and \cite{WuWeinbergerWelleniusDominiciBraun(21)}. 
Nonetheless, this is a complex causal effect as it measures an average causal effect of simultaneously shifting all assignments from time $t = 1$ to $t$.
If we only observe outcomes, separately identifying the causal effect of specific shocks on a particular outcome seems challenging without introducing additional structure on the potential outcome system.
    
\section{Conclusion}\label{sect:conclusion}
In this paper, we developed the direct potential outcome system to study causal inference in observational time series settings.
We place no functional form restrictions on the potential outcome process, no restrictions on the extent to which past assignments causally affect the outcomes, nor common time series assumptions such as ``invertibility' or ``recoverability.'' 
The direct potential outcome system therefore nests most leading econometric models used in time series settings as a special case.
We then studied conditions on the assignments under which common time series estimands, such as the impulse response functions, generalized impulse response function, and local projections, have a causal interpretation in terms of underlying dynamic causal effects. 
We further showed that provided the researcher observes an instrument that satisfies an appropriate unconfoundedness and monotonicity condition, then common IV estimands such as local projections instrumental variables also have causal interpretations in terms of local average dynamic causal effects. 
Taken together, the direct potential outcome system allows for making causal statements from observational time series, providing a foundation for studying the causal interpretation of reduced-form empirical strategies in time series settings. 

%%%%%%%%%%%%%%
% References %
%%%%%%%%%%%%%%
\baselineskip=12pt
\bibliographystyle{chicago}
\bibliography{Bibliography}
\baselineskip=20pt

%%%%%%%%%%%%
% Appendix %
%%%%%%%%%%%%
%\newpage \clearpage
\appendix
\section{Proofs of Results for Assignments and Outputs}

\subsection{Proof of Theorem \ref{theorem: IRF decomposition under full obs}}
To prove this result, we begin by rewriting $\mathbb{E}[ Y_{j, t+h} 1\{W_{k,t} = w_k\}]$. Notice that
    \[
    \mathbb{E}[ Y_{j, t+h} 1\{W_{k,t} = w_k\}]
    \]
    \[
    = \mathbb{E}[ Y_{j, t+h}(W_{1:t-1}, w_k, W_{-k,t}, W_{t+1:t+h}) 1\{W_{k,t} = w_k\} ] 
    \]
    \[
    = \mathbb{E}[ Y_{j, t+h}(W_{1:t-1}, w_k, W_{-k,t}, W_{t+1:t+h}) ] \mathbb{E}[ 1\{W_{k,t} = w_k\}]
    \] 
    \[
    + Cov\left( Y_{j, t+h}(W_{1:t-1}, w_k, W_{-k,t}, W_{t+1:t+h}), 1\{W_{k,t} = w_k\} \right).
    \]
Therefore, it immediately follows that 
    \[
     \mathbb{E}[ Y_{j, t+h} \mid W_{k,t} = w_k ] = \mathbb{E}[ Y_{j, t+h}(W_{1:t-1}, w_k, W_{-k,t}, W_{t+1:t+h}) ]
    \]
    \[
     + \frac{ Cov\left( Y_{j, t+h}(W_{1:t-1}, w_k, W_{-k,t}, W_{t+1:t+h}), 1\{W_{k,t} = w_k\} \right)}{ \mathbb{E}[ 1\{W_{k,t} = w_k\} ] }.
    \]
The result is then immediate by (i) applying the same calculation to $\mathbb{E}[ Y_{j, t+h} 1\{W_{k,t} = w_k^\prime\} ]$, (ii) taking the difference, and (iii) applying the definition of $Y_{j,t+h}(w_k)$. $\Box$

%subsection{Proof of Theorem \ref{theorem: shocks, IRFs}}
%The first claim is immediate. For the second claim, consider $w_{1:t+h} \in \mathcal{W}^{t+h}$ and notice that 
%    \[
%    P(Y_{j,t+h}(W_{1:t-1}, W_{1:k-1,t},w_{k,t}, W_{k+1:d_W,t}, W_{t+1:t+h}) = y_{j,t+h}, W_{1:t-1} = w_{1:t-1}, W_{-k,t} = w_{-k,t}, W_{t+1:t+h} = w_{t+1:t+h} \mid W_{k,t} = w_{k,t})
%    \]
%    \[
%    = P(Y_{j,t+h}(w_{1:t-1}, w_{k,t}, w_{-k,t}, w_{t+1:t+h}) = y_{j,t+h} \mid W_{1:t+h} = w_{1:t+h}) 
%    \]
%    \[
%    \times P(W_{1:t-1} = w_{1:t-1}, W_{-k,t} = w_{-k,t}, W_{t+1:t+h} = w_{t+1:t+h} \mid W_{k,t} = w_{k,t}) 
%    \]
%    \[
%    \overset{(1)}{=} P(Y_{j,t+h}(w_{1:t-1}, w_{k,t}, w_{-k,t}, w_{t+1:t+h}) = y_{j,t+h} \mid W_{1:t-1} = w_{1:t-1}, W_{-k,t} = w_{-k,t}, W_{t+1:t+h} = w_{t+1:t+h}) 
%    \]
%    \[
%    \times P(W_{1:t-1} = w_{1:t-1}, W_{-k,t} = w_{-k,t}, W_{t+1:t+h} = w_{t+1:t+h}) 
%    \]
%    \[
%    = P(Y_{j,t+h}(W_{1:t-1}, w_{k,t}, W_{-k,t}, W_{t+1:t+h}) = y_{j,t+h}, W_{1:t-1} = w_{1:t-1}, W_{-k,t} = w_{-k,t}, W_{t+1:t+h} = w_{t+1:t+h}),
%    \]
%where (1) follows because $W_{k,t} \indep (W_{1:t-1}, W_{-k,t}, W_{t+1:t+h}, \{Y_{t+h}(w_{1:t+h}) \colon w_{1:t+h} \in \mathcal{W}^{t+h}\})$ implies that $W_{k,t} \indep \{Y_{t+h}(w_{1:t+h}) \colon w_{1:t+h} \in \mathcal{W}^{t+h}\} \mid W_{1:t-1}, W_{-k,t}, W_{t+1:t+h}$. This proves the result since it shows that $W_{k,t} \indep \left( Y_{j,t+h}(w_k), W_{-k,t}, W_{t+1:t+h} \right)$, which implies that $W_{k,t} \indep Y_{j,t+h}(w_k)$. $\Box$

\subsection{Proof of Theorem \ref{theorem: local projection estimand, IRF}}  The style proof extends \cite{AngristGraddyImbens(00)-cts_late} in their analysis of the Wald estimand in a cross-sectional setting.  
Begin by writing $Y_{t+h} = Y_{t+h}(W_{k,t})$ as
    \begin{align*}
        Y_{t+h} &= Y_{t+h}(\underline{w}_k) + \int_{\underline{w}_k}^{W_{k,t}} \frac{\partial Y_{t+h}(\tilde{w}_k) }{d \tilde{w}_k } \partial\tilde{w}_k \\
        &= Y_{t+h}(\underline{w}_k) + \int_{\underline{w}_k}^{\overline{w}_{k}} \frac{\partial Y_{t+h}(\tilde{w}_k) }{\partial \tilde{w}_k } 1\{\tilde{w}_{k} \leq W_{k,t}\} d\tilde{w}_k
    \end{align*}
by the fundamental theorem of calculus. Then, it follows that
    \begin{align*}
        & Cov(Y_{t+h}, W_{k,t}) = \mathbb{E}[Y_{t+h} (W_{k,t} - \mathbb{E}[W_{k,t}]) ] \\
        &\overset{(1)}{=} \mathbb{E}[ ( Y_{t+h} - Y_{t+h}(\underline{w}_k) ) (W_{k,t} - \mathbb{E}[W_{k,t}]) \\
        &= \mathbb{E}\left[ \left( \int_{\underline{w}_k}^{\overline{w}_{k}} \frac{\partial Y_{t+h}(\tilde{w}_k) }{\partial \tilde{w}_k } 1\{\tilde{w}_{k} \leq W_{k,t}\} d\tilde{w}_k \right) (W_{k,t} - \mathbb{E}[W_{k,t}]) \right] \\
        &=  \int_{\underline{w}_k}^{\overline{w}_{k}} \mathbb{E}\left[ \frac{\partial Y_{t+h}(\tilde{w}_k) }{\partial \tilde{w}_k } 1\{\tilde{w}_{k} \leq W_{k,t}\} (W_{k,t} - \mathbb{E}[W_{k,t}]) \right] d\tilde{w}_k \\
        &\overset{(2)}{=} \int_{\underline{w}_k}^{\overline{w}_{k}} \mathbb{E}\left[ \frac{\partial Y_{t+h}(\tilde{w}_k) }{\partial \tilde{w}_k } \right] \mathbb{E}\left[ 1\{\tilde{w}_{k} \leq W_{k,t}\} (W_{k,t} - \mathbb{E}[W_{k,t}]) \right] d\tilde{w}_k
    \end{align*}
where (1) and (2) follow since $W_{k,t} \indep \{Y_{t+h}(w_k) \colon w_k \in \mathcal{W}_k\}$. Interchanging the order of the derivation and the expectation delivers the result.  Analogously, 
$$
W_{k,t} = \underline{w}_k + \int_{\underline{w}_k}^{W_{k,t}}  d\tilde{w}_k
= \underline{w}_k + \int_{\underline{w}_k}^{\overline{w}_{k}}   1\{\tilde{w}_{k} \leq W_{k,t}\} d\tilde{w}_k,
$$
so 
$$
Var(W_{k,t}) = \mathbb{E}[(W_{k,t}-\underline{w}_k)(W_{k,t} - \mathbb{E}[W_{k,t}])] = \int_{\underline{w}_k}^{\overline{w}_{k}} \mathbb{E}\left[ 1\{\tilde{w}_{k} \leq W_{k,t}\} (W_{k,t} - \mathbb{E}[W_{k,t}]) \right] d\tilde{w}_k.
$$
The result then follows immediately. To see that the resulting weights are non-negative, observe that for $\tilde{w}_k \in [\underline{w}_k, \overline{w}_k]$
    \[
    \mathbb{E}\left[ 1\{W_{k,t} \geq \tilde{w}_k\} \left( W_{k,t} - \mathbb{E}[W_{k,t}] \right) \right]
    \]
    \[
    = \mathbb{E}\left[ 1\{W_{k,t} \geq \tilde{w}_k\} W_{k,t} \right] - \mathbb{E}[1\{W_{k,t} \geq \tilde{w}_k\} ] \mathbb{E}[W_{k,t}] 
    \]
    \[
    = \left( \mathbb{E}\left[ W_{k,t} \mid W_{k,t} \geq \tilde{w}_k \right] - \mathbb{E}[W_{k,t}] \right) \mathbb{P}(W_{k,t} \geq \tilde{w}_k) \geq 0
    \]
since $\mathbb{E}\left[ W_{k,t} \mid W_{k,t} \geq \tilde{w}_k \right] \geq \mathbb{E}[W_{k,t}]$ for $\tilde{w}_k \in [\underline{w}_k, \overline{w}_k]$. $\Box$

\subsection{Proof of Theorem \ref{theorem: GIRF decomposition under full obs}}
The proof is analogous to the proof of Theorem \ref{theorem: IRF decomposition under full obs}. We start by rewriting $\mathbb{E}[ Y_{j, t+h} 1\{W_{k,t} = w_k\} \mid \mathcal{F}_{t-1}]$, noticing that
    \[
    \mathbb{E}[ Y_{j, t+h} 1\{W_{k,t} = w_k\} \mid \mathcal{F}_{t-1}]
    \]
    \[
    = \mathbb{E}[ Y_{j, t+h}(w_{1:t-1}^{obs}, w_k, W_{-k,t}, W_{t+1:t+h}) 1\{W_{k,t} = w_k\} \mid \mathcal{F}_{t-1} ] 
    \]
    \[
    = \mathbb{E}[ Y_{j, t+h}(w_{1:t-1}^{obs}, w_k, W_{-k,t}, W_{t+1:t+h}) \mid \mathcal{F}_{t-1} ] \mathbb{E}[ 1\{W_{k,t} = w_k\} \mid \mathcal{F}_{t-1} ]
    \] 
    \[
    + Cov\left( Y_{j, t+h}(w_{1:t-1}^{obs}, w_k, W_{-k,t}, W_{t+1:t+h}), 1\{W_{k,t} = w_k\} \mid \mathcal{F}_{t-1} \right).
    \]
Therefore, we have shown that
    \[
    \mathbb{E}[ Y_{j, t+h} \mid W_{k,t} = w_k, \mathcal{F}_{t-1} ] = \mathbb{E}[ Y_{j, t+h}(w_{1:t-1}^{obs}, w_k, W_{-k,t}, W_{t+1:t+h}) \mid \mathcal{F}_{t-1} ] 
    \]
    \[
    + \frac{ Cov\left( Y_{j, t+h}(w_{1:t-1}^{obs}, w_k, W_{-k,t}, W_{t+1:t+h}), 1\{W_{k,t} = w_k\} \mid \mathcal{F}_{t-1} \right)}{ \mathbb{E}[ 1\{W_{k,t} = w_k\} \mid \mathcal{F}_{t-1}] }.
    \]
The result follows by (i) applying the same calculation to $\mathbb{E}[ Y_{j, t+h} 1\{W_{k,t} = w_k^\prime\} \mid \mathcal{F}_{t-1}]$, (ii) taking the difference, and (iii) applying the definition of the potential outcome $Y_{j,t+h}(w_k)$. $\Box$

%\subsection{Proof of Theorem \ref{theorem: shocks, GIRFs}}
%The proof is the same as the Proof of Theorem \ref{theorem: shocks, IRFs}, except we must now condition on $\mathcal{F}_{t-1}$ throughout. $\Box$

\section{Proofs of Results for Assignments, Instruments and Outputs}

\subsection{Proof of Theorem \ref{thm:wald}}
To prove this result, we first observe that  
    \[
    \mathbb{E}[Y_{j,t+h} \mid Z_t = z] 
    = \mathbb{E}[Y_{j,t+h}(w_{1:t-1}^{obs}, W_{k,t}(z), W_{-k,t}, W_{t+1:t+h}) \mid Z_t = z]
    \]
    \[
    = \mathbb{E}[Y_{j,t+h}(w_{1:t-1}^{obs}, W_{k,t}(z), W_{-k,t}, W_{t+1:t+h})]
    \]
by (iii). Therefore, 
    \[
    \mathbb{E}[Y_{j,t+h} \mid Z_t = z] - \mathbb{E}[Y_{j,t+h} \mid Z_t = z^\prime]
    \]
    \[
    = \mathbb{E}[ Y_{j,t+h}(w_{1:t-1}^{obs}, W_{k,t}(z), W_{-k,t}, W_{t+1:t+h}) - Y_{j,t+h}(w_{1:t-1}^{obs}, W_{k,t}(z^\prime), W_{-k,t}, W_{t+1:t+h})].
    \]
Next, we can further rewrite this previous expression as
    \[
    \mathbb{E}[ \int_{W_{j,t}(z^\prime)}^{W_{j,t}(z)} \frac{\partial Y_{j,t+h}(w_k) }{\partial w_k} dw_k]
    = \mathbb{E}[ \int_{\cW}  \frac{\partial Y_{j,t+h}(w_k) }{\partial w_k} 1\{ W_{k,t}(z^\prime) \leq w_k \leq W_{k,t}(z) \} dw_k ] 
    \]
where we used the definition $Y_{j,t+h}(w_k) := Y_{j, t+h}(W_{1:t-1}, w_{k,t}, W_{-k, t}, W_{t+1:t+h})$. Finally, assuming that we can exchange the order of integration and expectation, we arrive at
    \begin{align*}
    &\int_{\cW} \mathbb{E}[ \frac{\partial Y_{j,t+h}(w_k) }{\partial w_k} 1\{ W_{k,t}(z^\prime) \leq w_k \leq W_{k,t}(z) \}] dw_k \\
    &=\int_{\cW} \mathbb{E}[ \frac{\partial Y_{j,t+h}(w_k) }{\partial w_k}, W_{k,t}(0) \leq w_k \leq W_{k,t}(1) ] \mathbb{E}[1\{W_{k,t}(z^\prime) \leq w_k \leq W_{k,t}(z)\}] dw_k.
    \end{align*}
We may apply the same argument to the denominator (again assuming that we can exchange the order of integration and expectation) to arrive at
    \[
    \mathbb{E}[W_{k,t} \mid Z_{t} = z] - \mathbb{E}[ W_{k, t} \mid Z_{t} = z^\prime] = 
    \]
    \[
    \mathbb{E}[W_{k,t}(z) - W_{k,t}(z^\prime)] = 
    \int_{\cW} \mathbb{E}[ 1\{W_{k,t}(z^\prime) \leq w_k \leq W_{k,t}(z)\}].
    \]
Taking the ratio then delivers the desired result. $\Box$

\subsection{Proof of Theorem \ref{theorem: instrument, observable treatments}} 

The proof is the same as the Proof of Theorem \ref{thm:wald}, except we must now condition on $\mathcal{F}_{t-1}$ throughout. $\Box$ 

\section{Proofs of Results for Outputs}

\subsection{Proof of Theorem \ref{thm:deterPO}}

%Write 
%$$
%\phi_{j,t+h}(w_{k,t}) = \mathrm{E}[Y_{j,t+h}(W_{k,t})|W_{k,t}=w_{k,t}]
%$$
Then, if the subsequent moments exist, we have that  
\begin{align*}
\mathrm{E}[Y_{j,t+h}|(Y_{k,t}=y_k),\mathcal{F}^Y_{t-1}] & = \mathrm{E}[Y_{j,t+h}(W_{1:t})|(Y_{k,t}=y_k),\mathcal{F}^Y_{t-1}],\quad \text{Assumption (i)}\\ 
&=\mathrm{E}[\mathrm{E}[Y_{j,t+h}(W_{1:t+h})|(Y_{k,t}=y_k),W_{1:t},\mathcal{F}^Y_{t-1}]|Y_{k,t}=y_k,\mathcal{F}^Y_{t-1}],\quad \text{Adam's law} \\ 
&=\mathrm{E}[\mathrm{E}[Y_{j,t+h}(W_{1:t+h})|W_{1:t})]|(Y_{k,t}=y_k),\mathcal{F}^Y_{t-1}],\quad \text{Assumption (i)} \\ 
%&=\mathrm{E}[\mathrm{E}[Y_{j,t+h}(W_{k,t})|W_{k,t}]|(Y_{k,t}=y_k),\mathcal{F}^Y_{t-1}],\quad %\text{Assumption \ref{ass:condInd}}, \\
& = \mathrm{E}[\psi_{j,t+h}(W_{1:t})|(Y_{k,t}=y_k),\mathcal{F}^Y_{t-1}], \quad \text{Assumption (ii)} 
\end{align*}
the last line holds as the future assignments are not informed by the historical ones. Applying this result twice gives the first result. $\Box$

\end{document}